\documentclass[11pt, letterpaper, onecolumn]{IEEEtran}

\usepackage{graphicx}
\usepackage{amsmath}
\usepackage{amssymb}
\usepackage{graphics}
\usepackage{latexsym}
\usepackage[dvips]{epsfig}
\usepackage{cite}
\usepackage{subfigure}
\usepackage{tabularx}
\usepackage{color}

\newtheorem{Definition}{Definition}
\newtheorem{Lemma}{Lemma}

\newtheorem{Proposition}[Lemma]{Proposition}
\newtheorem{Theorem}{Theorem}

\newtheorem{Remark}{Remark}

\def\Pr{{\mathrm{Pr}}}
\def\E{{\mathrm E}}
\def\Var{{\mathrm {Var}}}

\usepackage[ colorlinks = true,
 linkcolor = blue,
 urlcolor = blue,
 citecolor = red,
 anchorcolor = green,
]{hyperref}

\begin{document}
\title{A Proof of the Strong Converse Theorem for \\ Gaussian Multiple Access Channels}

\author{Silas~L.~Fong, \IEEEmembership{Member,~IEEE}  and Vincent~Y.~F.~Tan, \IEEEmembership{Senior Member,~IEEE}
\thanks{Silas~L.~Fong and Vincent~Y.~F.~Tan are with the Department of Electrical and Computer Engineering, National University of Singapore, Singapore (e-mail: \texttt{\{silas\_fong,vtan\}@nus.edu.sg}). Vincent~Y.~F.~Tan is also with the Department of Mathematics, National University of Singapore, Singapore. }}%
\maketitle
\flushbottom
\allowdisplaybreaks[1]
\begin{abstract}
We prove the strong converse for the $N$-source Gaussian multiple access channel (MAC). In particular, we show that any rate tuple that can be supported by a sequence of codes with asymptotic average error probability less than one must lie in the Cover-Wyner capacity region. Our proof consists of the following. First, we perform an expurgation step to convert any given sequence of codes with asymptotic average error probability less than one to codes with asymptotic maximal error probability less than one. Second, we quantize the input alphabets with an appropriately chosen resolution. Upon quantization, we apply the wringing technique (by Ahlswede) on the quantized inputs to obtain further subcodes from the subcodes obtained in the expurgation step so that the resultant correlations among the symbols transmitted by the different sources vanish as the blocklength grows.
Finally, we derive upper bounds on achievable sum-rates of the subcodes in terms of the type-II error of a binary hypothesis test. These upper bounds are then simplified through judicious choices of auxiliary output distributions. Our strong converse result carries over to the Gaussian interference channel under strong interference as long as the sum of the two asymptotic average error probabilities less than one.
\end{abstract}
\IEEEpeerreviewmaketitle

\section{Introduction}\label{introduction}
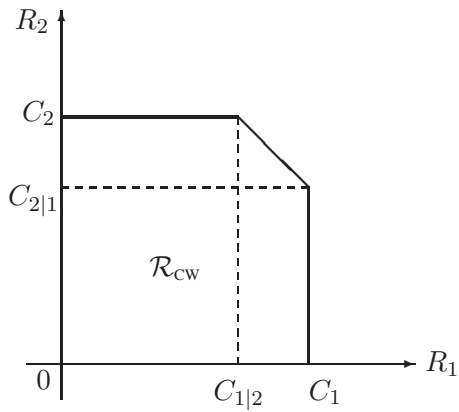
\begin{figure}
\centering
\begin{picture}(115, 135)
\setlength{\unitlength}{.47mm}
\multiput(10,60)(4,0){18}{\line(1,0){2}}
\multiput(60,10)(0,4){18}{\line(0,1){2}}
\put(0, 10){\vector(1, 0){110}}
\put(10, 0){\vector(0,1){110}}
\put(110,8){ $R_1$}
\put(-6, 105){ $R_2$}
\put(53, 0){$C_{1|2}$}
\put(80, 0){$C_1$}
\put(-5,55){$C_{2|1}$}
\put(-1,79){$C_2$}
\put(35,35){$\mathcal{R}_{\text{\tiny CW}}$}
\put(3,3){$0$}
\thicklines
\put(10, 80){\line(1, 0){50}}
\put(80, 10){\line(0,1){50}}
\put(60, 80){\line(1, -1){20}}
\end{picture}
\caption{Capacity region of the two-encoder Gaussian MAC \cite{cover_75,wyner74}. We use the shorthands $C_1 \triangleq \frac{1}{2}\log (1+P_1) $ and $C_{1|2} \triangleq\frac{1}{2}\log (1+P_1/(1+P_2))$ and similarly for $C_2$ and $C_{2|1}$. }
\label{fig:cap_reg}
\end{figure}

The multiple access channel (MAC) is one of the most well-studied problems in network information theory \cite{elgamal}. The capacity region of the discrete memoryless MAC was independently derived by Ahlswede \cite{ahl71} and Liao \cite{liao} in the early 1970s. In this paper, we are interested in the Gaussian version of this problem for which the channel output $Y$ corresponding to the inputs $(X_1, X_2,\ldots, X_N)$ is
\begin{equation}
Y=\sum_{i=1}^N X_i + Z,
\end{equation}
where $Z$ is standard Gaussian noise. We assume an average transmission power constraint of $P_i$ corresponding to each transmitter $i \in\{1,2,\ldots, N\}$. The capacity region was derived by Cover \cite{cover_75} and Wyner~\cite{wyner74} and is the set of all rate tuples $(R_1,R_2,\ldots, R_N) \in\mathbb{R}_+^N$ that satisfy
\begin{equation}
\sum_{i \in T} R_i \le \frac{1}{2} \log\bigg( 1+ \sum_{i\in T} P_i\bigg) \label{eqn:cover_wyner}
\end{equation}
for all subsets $T \subseteq \{1,2,\ldots, N\}$. For the $N=2$ case, the pentagonal region of rate tuples in \eqref{eqn:cover_wyner} is known as the {\em Cover-Wyner} region and is illustrated in Figure~\ref{fig:cap_reg}.

Despite our seemingly complete understanding of fundamental limits of the Gaussian MAC, it is worth highlighting that in the above-mentioned seminal works \cite{ahl71, liao, cover_75, wyner74}, it is assumed that the average error probability tends to zero as the length of the code grows without bound. This implies that those established converses are, in fact, {\em weak converses}. Fano's inequality~\cite[Sec.~2.1]{elgamal} is typically used as a key tool to establish such weak converses. In this work, we strengthen the results of Cover \cite{cover_75} and Wyner~\cite{wyner74} and show that any rate tuple that can be supported by a sequence (in the blocklength) of Gaussian multiple access codes with asymptotic average error probability {\em strictly less than one} (and not necessarily tending to zero) must lie in the Cover-Wyner region. This is a {\em strong converse} statement, akin to the work on strong converses for point-to-point channels by Wolfowitz~\cite{Wolfowitz}. It indicates that the boundary of the Cover-Wyner region designates a sharp phase transition of the smallest achievable asymptotic error probability, which is zero for any rate tuple inside the capacity region and one for any rate tuple outside the capacity region. Thus, this work augments our understanding of the first-order fundamental limit of the Gaussian MAC. Additionally, it may also serve as a stepping stone for studying the second-order asymptotics~\cite{Tan_FnT,Scarlett15, Mol13, TK14} or upper bounds (e.g., the sphere-packing bound) on the reliability function of the Gaussian MAC (cf.~\cite[Th.~4]{Weng08}).

\subsection{Related Work}
The study of MACs has a long history and we refer the reader to the excellent exposition in El Gamal and Kim \cite[Ch.~4]{elgamal} for a thorough discussion. Dueck \cite{dueck81} proved the strong converse for the (two-source) discrete memoryless MAC using the the technique of blowing up decoding sets originally due to Ahlswede, G\'acs and K\"orner \cite{Ahls76}, combined with a novel strategy known as the {\em wringing technique}. The technique of blowing up decoding sets uses the so-called {\em blowing-up lemma} \cite{Ahls76,Marton86} (see also \cite[Ch.~5]{Csi97} or \cite[Sec.~3.6]{RagSason}). This technique is useful for establishing strong converse results for memoryless channels with finite output alphabets.


Dueck's proof proceeds in three steps. First, Dueck expurgates an appropriate subset of  codeword pairs to convert any given sequence of codes with asymptotic \emph{average} error probability bounded away from one to a sequence of codes with asymptotic \emph{maximal} error probability bounded away from one.\footnote{Although the capacity region of the Gaussian MAC is well-known when it is defined in terms of the average error probability \cite{elgamal,cover_75,wyner74}, the determination of the capacity region is an open problem if it is defined in terms of the {\em maximal} error probability. } This expurgation step is performed so that the blowing-up lemma to be applied in the third step yields tight upper bounds on the sum-rate, which will then lead to the desired strong converse result.  Unfortunately, the expurgation step introduces undesirable correlations among the codewords transmitted by the~$N$ encoders. Second, a wringing technique is introduced to wring out any residual dependence among the symbols transmitted by the $N$ encoders by choosing a further subcode from each subcode obtained in the expurgation step. Wringing is necessary for establishing a tight sum-rate bound, because the sum-rate capacity of the MAC is expressed as the supremum of mutual information terms over all independent input distributions (the independence is due to the fact that the $N$ encoders do not cooperate). Third, the blowing-up lemma is applied to the resultant subcode to yield a tight upper bound on the sum-rate.

Ahlswede \cite{Ahl82} presented another strong converse proof for the (two-source) discrete memoryless MAC by modifying Dueck's wringing technique as well as  replacing the use of the blowing-up lemma in Dueck's proof with an application of Augustin's non-asymptotic converse bound~\cite{augustin}. However, the proofs of Dueck and Ahlswede are specific to the discrete (finite alphabet) setting and it is not clear by examining the proofs that the same strong converse statement follows in a straightforward way for the Gaussian MAC with peak power constraints.

Another approach to proving the strong converse for a general MAC is due to Han \cite{Han98}, who used the information spectrum technique~\cite{Han10} to provide a general formula for MACs and stated a condition \cite[Th.~6]{Han98} for the strong converse to hold. However, unlike for the point-to-point setting~\cite[Sec.~3.6--3.7]{Han10}, the property is difficult to verify for various classes of memoryless MACs.

In view of the above works and the practical  and theoretical importance of strong converse theorems, we are motivated to provide a self-contained proof for the strong converse of the Gaussian MAC.

\subsection{Challenges in Establishing the Strong Converse and Our Strategies to Overcome Them} \label{sectionChallenges}
In this subsection, we discuss the challenges of leveraging existing techniques to prove the strong converse for the Gaussian MAC. In particular, we highlight the difficulties in directly using the ideas contained in Dueck's \cite{dueck81} and Ahlswede's~\cite{Ahl82} proofs. We also describe, at a high level, the strategy we employ to overcome these difficulties. Finally, we discuss some other auxiliary proof techniques.

\subsubsection{Blowing-Up Lemma in Dueck's Proof \cite{dueck81} Cannot be Directly Extended to Continuous Alphabets}\label{sec:blowingUp}
In Dueck's paper~\cite{dueck81}, he used a version of the {\em blowing-up lemma}, together with other tools, to prove the strong converse    theorem for the discrete memoryless MAC.  A crucial step in Dueck's proof involves the establishing of an upper bound on the list size of possible messages for every output sequence based on the blown-up decoding sets. If the resultant list size is too large (e.g., contains an exponential number of messages),  the Dueck's technique cannot lead to the strong converse theorem. Since this crucial step heavily relies on the finiteness of the output alphabet and the output alphabet of the Gaussian MAC is uncountably infinite, it is not immediately apparent how to extend this step to the Gaussian case.

\subsubsection{Wringing Technique in Ahlswede's Proof \cite{Ahl82} Cannot be Directly Extended to Continuous Alphabets} \label{sec:wringing}
As mentioned in the previous section, Ahlswede's proof \cite{Ahl82} is based on a modification of Dueck's wringing technique and Augustin's non-asymptotic converse bound~\cite{augustin}.
  However, it is not apparent how to adapt his techniques to obtain a strong converse bound on the sum-rate. More specifically, Ahlswede's wringing technique (see Equation (5.3) in \cite{Ahl82}) leads to the following sum-rate bound for any sequence of length-$n$ codes whose asymptotic average error probability is bounded away from one:
\begin{equation}
R_{1}  + R_{2} \le I(X_1, X_2; Y) + O\left(\frac{\log n}{\sqrt{n}}\right) |\mathcal{X}_1||\mathcal{X}_2||\mathcal{Y}| . \label{eqnAhlswede}
\end{equation}
In \eqref{eqnAhlswede}, $X_1$ and $X_2$ are  {\em independent} random variables. However, the bound in \eqref{eqnAhlswede} is sensitive to the sizes of the input and output alphabets, which prevents us from directly extending Ahlswede's proof to the Gaussian case. Furthermore, there are no cost constraints in the discrete memoryless MAC and incorporating cost constraints does not seem to be  trivial. A na\"ive strategy to extend Ahlswede's proof to the Gaussian case is to quantize the input and output alphabets of the Gaussian MAC so that $\mathcal{X}_{1}$, $\mathcal{X}_{2}$  and $\mathcal{Y}$ depend on $n$ and their cardinalities grow with~$n$. Say we denote the quantized alphabets as $\hat{\mathcal{X}}_{1}^{ (n) }$, $\hat{\mathcal{X}}_{2}^{ (n) }$  and $\hat{\mathcal{Y}}^{ (n) }$. This sequence of quantized alphabets and the corresponding channels will be designed to provide increasingly refined approximations to the Gaussian MAC as $n$ increases.  In designing $\hat{\mathcal{X}}_{1}^{ (n) }$, $\hat{\mathcal{X}}_{2}^{ (n) }$  and $\hat{\mathcal{Y}}^{ (n) }$, we would also like to ensure that  the power constraints are satisfied and  the term $O\left(\frac{\log n}{\sqrt{n}}\right)| \hat{\mathcal{X}}_{1}^{ (n) }||\hat{\mathcal{X}}_{2}^{ (n) }||\hat{\mathcal{Y}}^{ (n) }|$ in \eqref{eqnAhlswede} vanishes as~$n$ tends to infinity. However, quantization arguments that are used to prove information-theoretic statements for continuous-valued alphabets are usually applied to the achievability parts of coding theorems. For example, a quantization argument is used in \cite[Sec.~3.4.1]{elgamal} for leveraging the achievability proof for the discrete memoryless channel (DMC) with cost constraints to prove the achievability part of the capacity of the AWGN channel.  To the best of our knowledge, standard quantization arguments for achievability parts do not work for strong converse proofs because upon quantization, one has to ensure that the resultant asymptotic error probability is bounded away from one.

The reader is also referred to \cite[Appendix~D.6]{MolavianJaziThesis} for a complementary explanation of why Ahlswede's original wringing technique works for only MACs with finite alphabets but not the Gaussian MAC.

\subsubsection{Remedy -- Combining a Quantization Argument with the Wringing Technique}  \label{sec:remedy}
The difficulties in directly using Dueck's and Ahlswede's techniques led the authors to combine a novel quantization argument together with Ahlswede's wringing idea. We use a scalar quantizer of increasing precision in the blocklength to discretize (only) the input alphabets of the channel so that the Ahlswede's wringing technique can be performed {\em on the quantized channel inputs} for any given code whose asymptotic error probability is bounded away from one. In doing so, we obtain a sequence of subcodes whose asymptotic error probability is bounded away from one such that the resultant correlations among the codeword symbols transmitted by the different sources vanish as~$n$ increases.
Note that if the quantizer's precision is too small or too large, the resultant upper bound on the sum-rate will be too loose and hence not useful in proving the strong converse. We discuss feasible choices of the quantizer's precision and the parameters used in the wringing technique in Section~\ref{sectionDiscussionQuantizer}. In our proof, the quantizer's precision is chosen in such a way that the quantized input alphabets $\hat{\mathcal{X}}_i^{(n)}$ grow no faster than $O(n^{3/2})$. It turns out that this choice of quantization also allows us to  control  the approximation errors between the true channel inputs and the quantized ones uniformly.

\subsubsection{Other Ingredients in Our Proof} \label{sec:otherIngredients}
In Ahlswede's proof of the strong converse for the discrete memoryless MAC, he appealed to a non-asymptotic converse bound by Augustin~\cite{augustin}. In our proof we use a conceptually similar non-asymptotic converse bound that is motivated by modern techniques  relating binary hypothesis testing to channel coding. In particular, we use a form of the {\em meta-converse} \cite[Sec.~III-E]{PPV10}  due to  Wang, Colbeck and Renner~\cite[Lemma~1]{wang09}. We derive a multi-user version of this  non-asymptotic converse bound. After doing so, we choose the auxiliary conditional output distributions therein to be product distributions that approximate  the  {\em quantized} code distribution.  We note that the flexibility of the choice of the output distributions is essential for proving the strong converse for the Gaussian MAC as we can allow these distributions to depend not only on the peak powers but also the chosen precision of the scalar quantizer (cf.\ Section~\ref{sec:remedy}).

\subsection{Paper Outline}
In the next subsection, we state the notation used in this paper. In Section \ref{sectionDefinition}, we describe the system model and define the $\varepsilon$-capacity region of the Gaussian MAC. In Section \ref{sec:main_res}, we present the main result of the paper. We present a few preliminaries for the proof in Section \ref{sectionPrelim}. The complete proof is then presented in Section \ref{sec:prf_main_result}. Section~\ref{sectionIC} extends our strong converse result to the two-source two-destination Gaussian IC under strong interference.
\subsection{Notation}\label{notation}
We use the upper case letter~$X$ to denote an arbitrary (discrete or continuous) random variable with alphabet $\mathcal{X}$, and use a lower case letter $x$ to denote a realization of~$X$.
We use $X^n$ to denote the random tuple $(X_1, X_2, \ldots, X_n)$. 

The following notations are used for any arbitrary random variables~$X$ and~$Y$ and any mapping $g$ whose domain includes $\mathcal{X}$. We let $p_{X,Y}$ and $p_{Y|X}$ denote the probability distribution of $(X,Y)$ (can be both discrete, both continuous or one discrete and one continuous) and the conditional probability distribution of $Y$ given $X$ respectively.
We let $p_{X,Y}(x,y)$ and $p_{Y|X}(y|x)$ be the evaluations of $p_{X,Y}$ and $p_{Y|X}$ respectively at $(X,Y)=(x,y)$. To avoid confusion, we do not write $\Pr\{X=x, Y=y\}$ to represent $p_{X,Y}(x,y)$ unless $X$ and $Y$ are both discrete.
To make the dependence on the distribution explicit, we let $\Pr_{p_X}\{ g(X)\in\mathcal{A}\}$ denote $\int_{x\in \mathcal{X}} p_X(x)\mathbf{1}\{g(x)\in\mathcal{A}\}\, \mathrm{d}x$ for any real-valued function~$g$ and any set $\mathcal{A}$.
The expectation and the variance of~$g(X)$ are denoted as
$
\E_{p_X}[g(X)]$ and
$
 \Var_{p_X}[g(X)]=\E_{p_X}[(g(X)-\E_{p_X}[g(X)])^2]$
 respectively, where we again make the dependence on the underlying distribution $p_X$ explicit.
 We let $\mathcal{N}(\,\cdot\, ;\mu,\sigma^2): \mathbb{R}\rightarrow [0,\infty)$ denote the probability density function of a Gaussian random variable whose mean and variance are $\mu$ and $\sigma^2$ respectively. This means that
\begin{equation}
\mathcal{N}(z;\mu,\sigma^2)\triangleq\frac{1}{\sqrt{2\pi \sigma^2}}\exp\bigg(-\frac{(z-\mu)^2}{2\sigma^2} \bigg).
\end{equation}
 We will take all logarithms to base 2 throughout this paper.
 The Euclidean norm of a vector $x^n\in\mathbb{R}^n$ is denoted by $\|x^n\|=\sqrt{\sum_{k=1}^n x_k^2}$.

\section{Gaussian Multiple Access Channel} \label{sectionDefinition}
We consider a Gaussian MAC that consists of $N$ sources and one destination. Let
\begin{equation}
\mathcal{I}\triangleq \{1, 2, \ldots, N\}
\end{equation}
 be the index set of the sources (or encoders), and let $\mathrm{d}$ denote the destination (or decoder). The $N$ message sources transmit information to the destination in $n$ time slots (channel uses) as follows. For each $i\in \mathcal{I}$, node~$i$ chooses message
\begin{equation}
W_{i}\in \{1, 2, \ldots, M_i^{(n)}\}
\end{equation}
 and sends $W_{i}$ to node~$\mathrm{d}$ where $M_i^{(n)}$ denotes the message size.
 Based on $W_i$, each node~$i$ prepares a codeword $X_i^n\in \mathbb{R}^n$ to be transmitted and $X_i^n$ should satisfy
 \[\sum_{k=1}^nX_{i,k}^2 \le n P_i,\] where $P_i$ denotes the power constraint for the codeword transmitted by node~$i$. Then for each $k\in \{1, 2, \ldots, n\}$, each node~$i$ transmits $X_{i,k}$ in time slot~$k$ and node~$\mathrm{d}$ receives the real-valued symbol
 \begin{equation}
 Y_{k}=\sum_{i\in \mathcal{I}} X_{i,k}+Z_k,
\end{equation}
 where $Z_1, Z_2, \ldots, Z_n$ are i.i.d.\ and $Z_1$ is a standard Gaussian random variable. After~$n$ time slots, node~$\mathrm{d}$ declares~$\{\hat W_i\}_{i\in\mathcal{I}}$ to be the transmitted~$\{W_i\}_{i\in\mathcal{I}}$ based on $Y^n$.

To simplify notation, we use the following convention for any $T\subseteq \mathcal{I}$. For any random tuple $(X_{1}, X_{2}, \ldots, X_N)$, we let
 \begin{equation}
 X_T\triangleq (X_{i} \,|\, i\in T)
\end{equation}
be its subtuple, whose generic realization and alphabet are denoted by $x_T$ and
\begin{equation}
\mathcal{X}_T =\prod_{i\in T} \mathcal{X}_i
\end{equation}
 respectively.
Similarly, for any $k\in \{1, 2, \ldots, n\}$ and any random tuple $(X_{1,k}, X_{2,k}, \ldots, X_{N,k})\in \mathcal{X}_\mathcal{I}$, we let
 \begin{equation}
X_{T,k}\triangleq (X_{i,k} \,|\, i\in T)
\end{equation}
 be its subtuple, whose realization is denoted by $x_{T,k}$.
The following five definitions formally define a Gaussian MAC and its capacity region.
\medskip
\begin{Definition} \label{defCode}
Let $T$ be a non-empty subset in $\mathcal{I}$. An {\em $(n, M_\mathcal{I}^{(n)}, P_\mathcal{I}, \mathcal{A}, T)$-code} for the Gaussian MAC, where $M_\mathcal{I}^{(n)}\triangleq (M_1^{(n)}, M_2^{(n)}, \ldots, M_N^{(n)})$ and $P_\mathcal{I}\triangleq (P_1, P_2, \ldots, P_N)$, consists of the
following:
\begin{enumerate}
\item A message set
$
\mathcal{W}_{i}\triangleq \{1, 2, \ldots, M_i^{(n)}\}
$
 at node~$i$ for each $i\in \mathcal{I}$.

 \item A support set of the message tuple $W_\mathcal{I}$ denoted by $\mathcal{A}\subseteq \mathcal{W}_\mathcal{I}$ where $W_\mathcal{I}$ is uniform on $\mathcal{A}$. In addition, all the $w_\mathcal{I}$'s in $\mathcal{A}$ have the same $w_{T^c}$, i.e., there exists a $w_{T^c}^*\in \mathcal{W}_{T^c}$ such that for all $w_\mathcal{I}\in \mathcal{A}$, we have $w_{T^c}=w_{T^c}^*$. Define
 \begin{equation}
 \mathcal{A}_T\triangleq\{w_T\in \mathcal{W}_T\,|\,\text{There exists a }\tilde w_{\mathcal{I}}\in\mathcal{A} \text{ such that }w_T=\tilde w_T\} \label{defSetAT}
 \end{equation}
 to be the support of $W_T$. Consequently, the message tuple $W_T$ is uniform on $\mathcal{A}_T$.

\item An encoding function
$
f_i : \mathcal{W}_{i}\rightarrow \mathbb{R}^n
$
 for each $i\in \mathcal{I}$, where $f_i$ is the encoding function at node~$i$ such that
$
X_{i}^n=f_i(W_{i})
$
and
\begin{equation}
\|f_i(w_{i})\|^2 \le nP_i \label{powerConstraint}
\end{equation}
for all $w_i\in\mathcal{W}_i$. The set of codewords $\{f_i(1), f_i(2), \ldots, f_i(M_i^{(n)})\}$ is called the \textit{codebook for $W_i$}. For each $i\in\mathcal{I}$, the finite alphabet
\begin{equation}
\mathcal{X}_i \triangleq \left\{x\in \mathbb{R}\left| \text{$x$ is a component of } f_i(w_i) \text{ for some }w_i \in \mathcal{W}_i \right.\right\} \label{defAlphabetXi*}
\end{equation}
is called the \textit{support} of symbols transmitted by~$i$ because $f_i(\mathcal{W}_i)\subseteq \mathcal{X}_i^n$. Note that
\begin{equation}
|\mathcal{X}_i| \le nM_i^{(n)} \label{defAlphabetXi}
\end{equation}
 for each $i\in \mathcal{I}$ by \eqref{defAlphabetXi*}.
\item A (possibly stochastic) decoding function
$
\varphi:
\mathbb{R}^{n} \rightarrow \mathcal{A},
$
which is used by node~$\mathrm{d}$ to estimate the message tuple $W_{\mathcal{I}}$, i.e., $
 \hat W_{\mathcal{I}} = \varphi(Y^{n})$.
\end{enumerate}
\end{Definition}
\smallskip
If $\mathcal{A}=\mathcal{W}_{\mathcal{I}}$ and $T=\mathcal{I}$, then $W_\mathcal{I}$ is uniformly distributed on $\mathcal{W}_{\mathcal{I}}$, which implies that the $N$ messages are mutually independent. Since $(n, M_\mathcal{I}^{(n)}, P_\mathcal{I}, \mathcal{W}_\mathcal{I}, \mathcal{I})$-codes are of our main interest, they are also called $(n, M_\mathcal{I}^{(n)}, P_\mathcal{I})$-codes for notational convenience. However, in the present work, it is necessary to allow $\mathcal{A}$ and $T$ to be strict subsets of $\mathcal{W}_{\mathcal{I}}$ and $\mathcal{I}$ respectively so the generality afforded in the above definition is necessary. In this case, the $N$ messages need not be independent.
In the rest of this paper, if we fix a code with encoding functions $\{f_i|i\in\mathcal{I}\}$, then $\mathcal{X}_i$ as defined in~\eqref{defAlphabetXi*} denotes the support of symbols transmitted by each $i\in\mathcal{I}$. \smallskip
\medskip
\begin{Definition}\label{defGaussianMAC}
A {\em Gaussian MAC} is characterized by the conditional probability density function
$q_{Y|X_\mathcal{I}}$ satisfying
\begin{equation}
q_{Y|X_\mathcal{I}}(y|x_\mathcal{I}) = \mathcal{N} \left(y ; \sum_{i\in\mathcal{I}} x_i, 1\right) \label{defChannelInDefinition}
\end{equation}
for all $x_\mathcal{I}\in \mathbb{R}^N$ and all $y\in\mathbb{R}$ such that the following holds for any $(n, M_{\mathcal{I}}^{(n)}, P_\mathcal{I}, \mathcal{A}, T)$-code: Let $p_{W_{\mathcal{I}} , X_{\mathcal{I}}^n , Y^n}$ be the probability distribution induced by the $(n, M_{\mathcal{I}}^{(n)}, P_\mathcal{I}, \mathcal{A}, T)$-code. Then,
\begin{align}
p_{W_{\mathcal{I}} , X_{\mathcal{I}}^n , Y^n}(w_{\mathcal{I}}, x_{\mathcal{I}}^n, y^n) = p_{W_{\mathcal{I}}}(w_\mathcal{I}) \left( \prod_{i\in\mathcal{I}}\mathbf{1}\{ x_{i}^n=f_i(w_i)\} \right)\left(\prod_{k=1}^n p_{Y_k|X_{\mathcal{I},k}}(y_k | x_{\mathcal{I}, k}) \right)\label{memorylessStatement*}
\end{align}
for all $(w_\mathcal{I}, x_\mathcal{I}^n, y^n)\in \mathcal{A}\times \mathcal{X}_{\mathcal{I}}^n\times \mathbb{R}^n$ where
\begin{equation}
p_{Y_k|X_{\mathcal{I},k}}(y_k |x_{\mathcal{I},k})\triangleq q_{Y|X_{\mathcal{I}}}(y_k | x_{\mathcal{I},k}). \label{defChannelInDefinition*}
\end{equation}
Since $p_{Y_k|X_{\mathcal{I},k}}$ does not depend on~$k$ by \eqref{defChannelInDefinition*} and \eqref{defChannelInDefinition}, the channel is stationary.
\end{Definition}
\medskip

For any $(n, M_{\mathcal{I}}^{(n)}, P_\mathcal{I}, \mathcal{A}, T)$-code defined on the Gaussian MAC, let $p_{W_{\mathcal{I}},X_\mathcal{I}^n, Y^n, \hat W_{\mathcal{I}}}$ be the joint distribution induced by the code. Since $\hat W_\mathcal{I}$ is a function of $Y^n$ by Definition~\ref{defCode}, it follows that
\begin{equation}
p_{W_{\mathcal{I}},X_\mathcal{I}^n, Y^n, \hat W_{\mathcal{I}}}=p_{W_{\mathcal{I}},X_\mathcal{I}^n, Y^n}p_{\hat W_\mathcal{I} |Y^n},
\end{equation}
which implies from \eqref{memorylessStatement*} that
\begin{equation}
p_{W_{\mathcal{I}},X_\mathcal{I}^n, Y^n, \hat W_{\mathcal{I}}}
=p_{W_{\mathcal{I}},X_\mathcal{I}^n}\left(\prod_{k=1}^n p_{Y_k|X_{\mathcal{I},k}}\right)p_{\hat W_\mathcal{I} |Y^n}. \label{memorylessStatement}
\end{equation}
\medskip
\begin{Definition} \label{defError}
For an $(n, M_{\mathcal{I}}^{(n)}, P_\mathcal{I})$-code defined on the Gaussian MAC, we can calculate according to \eqref{memorylessStatement} the {\em average probability of decoding error} which is defined as
\begin{equation}
\Pr\big\{\hat W_{\mathcal{I}} \ne W_{\mathcal{I}}\big\}.
\end{equation}
An $(n, M_{\mathcal{I}}^{(n)}, P_\mathcal{I})$-code with average probability of decoding error no larger than $\varepsilon$ is called an $(n, M_{\mathcal{I}}^{(n)}, P_\mathcal{I}, \varepsilon)_{\text{avg}}$-code.
Similarly for an $(n, M_{\mathcal{I}}^{(n)}, P_\mathcal{I}, \mathcal{A}, T)$-code, we can calculate the {\em maximal probability of decoding error} defined as
\begin{equation}
\max_{w_\mathcal{I}\in \mathcal{A}}\Pr\big\{\hat W_T\ne W_T \, \big|\, W_\mathcal{I}=w_\mathcal{I} \big\}.
\end{equation}
An $(n, M_{\mathcal{I}}^{(n)}, P_\mathcal{I}, \mathcal{A}, T)$-code with maximal probability of decoding error no larger than $\varepsilon$ is called an $(n, M_{\mathcal{I}}^{(n)}, P_\mathcal{I}, \linebreak \mathcal{A}, T, \varepsilon)_{\text{max}}$-code.
\end{Definition}
\medskip
\begin{Definition} \label{defAchievableRate}
A rate tuple $R_\mathcal{I}\triangleq (R_1, R_2, \ldots, R_N)$ is \textit{$\varepsilon$-achievable} for the Gaussian MAC if there exists a sequence of $(n, M_{\mathcal{I}}^{(n)}, P_\mathcal{I}, \varepsilon_n)_{\text{avg}}$-codes on the Gaussian MAC such that
\begin{equation}
\liminf\limits_{n\rightarrow \infty}\frac{1}{n}\log M_i^{(n)} \ge R_i
\end{equation}
for each $i\in\mathcal{I}$ and
\begin{equation}
\limsup\limits_{n\rightarrow \infty}\varepsilon_n \le \varepsilon.\label{eqn:asymp_err}
\end{equation}
\end{Definition}
\medskip
\begin{Definition}\label{defCapacityRegion}
For each $\varepsilon \in [0,1)$, the \textit{$\varepsilon$-capacity region} of the Gaussian MAC, denoted by $\mathcal{C}_\varepsilon$, is the set consisting of all $\varepsilon$-achievable rate tuples $R_{\mathcal{I}}$. The \textit{capacity region} is defined to be the $0$-capacity region $\mathcal{C}_0$.
\end{Definition}


\section{Main Result} \label{sec:main_res}
The following theorem is the main result in this paper.

\begin{Theorem} \label{thmMainResult}
Define
\begin{equation}
\mathcal{R}_{\text{\tiny CW}} \triangleq \bigcap_{T\subseteq \mathcal{I}} \left\{R_\mathcal{I} \in \mathbb{R}_+^N \left| \: \parbox[c]{2 in}{$\sum_{i\in T}R_i\le \frac{1}{2}\log\big(1+\sum_{i\in T}P_i\big)$} \right.\right\}. \label{Rout}
\end{equation}
 Then for each $\varepsilon \in [0,1)$,
\begin{equation}
\mathcal{C}_\varepsilon \subseteq \mathcal{R}_{\text{\tiny CW}}.
\end{equation}
\end{Theorem}

We now present three remarks concerning Theorem \ref{thmMainResult}.
\begin{enumerate}

\item Note that $\mathcal{R}_{\text{\tiny CW}} $ is the Cover-Wyner \cite{cover_75, wyner74} region for an $N$-source Gaussian MAC. The theorem says that regardless of the admissible average error probability (as long as it is strictly smaller than $1$), all achievable rate tuples must lie in $\mathcal{R}_{\text{\tiny CW}} $. Since all rate tuples in $\mathcal{R}_{\text{\tiny CW}} $ are $0$-achievable~\cite[Sec.~4.7]{elgamal},    we have for every $\varepsilon\in [0,1)$
\begin{equation}
\mathcal{C}_\varepsilon =\mathcal{R}_{\text{\tiny CW}}.
\end{equation}

\item In fact, the proof allows us to additionally assert the following: For any non-vanishing average error probability $\varepsilon\in [0,1)$ and any subset $T\subseteq \mathcal{I}$, it can be shown that the sum-rate of the messages indexed by $T$ of any sequence of $(n, M_{\mathcal{I}}^{(n)},P_{\mathcal{I}},\varepsilon_n)_{\text{avg}}$-codes satisfying the constraint in~\eqref{eqn:asymp_err}
 also satisfies
\begin{equation}
\limsup_{n\to\infty}\frac{1}{\sqrt{n\log n}} \left[ \sum_{i\in T}\log M_i^{(n)} - \frac{ n}{2}\log\bigg(1+ \sum_{i\in T} P_i\bigg) \right] \le\overline{\Upsilon}(\varepsilon,T , P_{\mathcal{I}})<\infty
\label{eqn:sec_order}
\end{equation}
for some finite constant $\overline{\Upsilon}(\varepsilon,T , P_{\mathcal{I}})$.
See \eqref{eqnBHT17thChain} in the proof of Theorem \ref{thmMainResult}. Even though the normalizing speed of $\sqrt{n \log n}$ is not the desired $\sqrt{n }$ (as usually defined in second-order asymptotic analyses~\cite{Tan_FnT}), the techniques in this work may serve as a stepping stone to establish an outer bound for the second-order coding rate region \cite{Tan_FnT} for the Gaussian MAC. The best inner bound for the second-order coding rates for the Gaussian MAC was established independently by Scarlett, Martinez, and Guill\'en i F\`abregas~\cite{Scarlett15} and MolavianJazi and Laneman~\cite{Mol13}. According to the inner bounds in~\cite{Scarlett15, Mol13} and the relation between second-order coding rates and second-order asymptotics of sum-rates in~\cite{TK14},
\begin{equation}
\liminf_{n\to\infty}\frac{1}{\sqrt{n}} \left[ \sum_{i\in T}\log M_i^{(n)} - \frac{ n}{2}\log\bigg(1+ \sum_{i\in T} P_i\bigg) \right] \ge\underline{\Upsilon}(\varepsilon, T, P_{\mathcal{I}})> -\infty \label{eqn:sec_inner}
\end{equation}
for some finite constant $\underline{\Upsilon}(\varepsilon, T, P_{\mathcal{I}})$.
Our normalizing speed of $\sqrt{n\log n }$ in \eqref{eqn:sec_order} is slightly better than in Ahlswede's work on the discrete memoryless MAC~\cite{Ahl82}, which is $\sqrt{n} \log n$. We have attempted to optimize (reduce) the  exponent of the logarithm $\zeta>0$ in the normalizing speed $\sqrt{n}(\log n)^\zeta$. However, as we will discuss in Section~\ref{sectionDiscussionQuantizer} in the sequel, we are unable to use our proof technique to further reduce (improve) $\zeta$ from~$1/2$. For both the discrete and Gaussian MACs, it is  challenging to prove that the exact normalizing speed of the second-order term is $\sqrt{n}$. This is, in part,  due to the use of wringing technique in the converse part, which prevents one from obtaining a converse that matches the achievability in the rate of growth of the second-order term.
Unless new techniques are invented to replace the wringing argument in the strong converse proof for the MAC (such techniques have remained elusive for over 30 years), the exact normalizing speed of the second-order term for the discrete and Gaussian MACs will remain an open problem.   

\end{enumerate}
 In the next section, we will present a few preliminaries for the proof of Theorem~\ref{thmMainResult}, which will be detailed in Section \ref{sec:prf_main_result}.
\section{Preliminaries for the Proof of Theorem \ref{thmMainResult}} \label{sectionPrelim}
\subsection{Expurgation of Message Tuples} \label{sectionExpurgation}
The following lemma is based on the technique of expurgating message tuples introduced by Dueck~\cite[Sec.~II]{dueck81}, and the proof is provided in the Appendix for completeness.
 \medskip
\begin{Lemma}\label{lemmaExpurgation}
Let $\varepsilon\in[0,1)$. Suppose an $(n, M_{\mathcal{I}}^{(n)}, P_\mathcal{I}, \varepsilon)_{\text{avg}}$-code for the Gaussian MAC is given. Then for each nonempty $T\subseteq \mathcal{I}$ such that
\begin{equation}
\left\lfloor \left(\frac{1-\varepsilon}{1+\varepsilon}\right)\prod_{i\in T}M_i^{(n)} \right\rfloor \ge \left(\frac{1-\varepsilon}{2(1+\varepsilon)}\right)\prod_{i\in T}M_i^{(n)}, \label{assumptionInLemmaWringing}
\end{equation}
there exist a set $\mathcal{A}\subseteq \mathcal{W}_\mathcal{I}$ and an $\big(n, M_{\mathcal{I}}^{(n)}, P_\mathcal{I}, \mathcal{A}, T , \frac{1+\varepsilon}{2}\big)_{\text{max}}$-code such that
\begin{equation}
|\mathcal{A}_T| = |\mathcal{A}| \ge \left(\frac{1-\varepsilon}{2(1+\varepsilon)}\right)\prod_{i\in T}M_i^{(n)}, \label{lemmaWringingSt}
\end{equation}
where $\mathcal{A}_T$ is as defined in \eqref{defSetAT}. As a consequence, if we let $p_{W_\mathcal{I}, X_\mathcal{I}^n, Y^n, \hat W_\mathcal{I}}$
  denote the probability distribution induced on the Gaussian MAC by the $\big(n, M_{\mathcal{I}}^{(n)}, P_\mathcal{I}, \mathcal{A}, T, \frac{1+\varepsilon}{2}\big)_{\text{max}}$-code, then we have for each $w_T\in \mathcal{A}_T$
\begin{equation}
p_{W_T}(w_T) \le \frac{1}{\prod_{i\in T}M_i^{(n)}}\cdot\left(\frac{2(1+\varepsilon)}{1-\varepsilon}\right).
 \label{st2LemmaWringing}
\end{equation}
\end{Lemma}
\begin{Remark}
Lemma~\ref{lemmaExpurgation} says that restricted to the set $\mathcal{A}_T$, the $i^{\text{th}}$ (for $i\in T$) codebooks have almost the same sizes as the original codebooks. In addition, the conditional probability of decoding error for each message tuple in this restricted codebook is upper bounded by $\frac{1+\varepsilon}{2}$, which is still smaller than one because $\varepsilon \in [0,1)$. According to~\eqref{st2LemmaWringing}, the probability of each message tuple cannot be greater than its original value by a factor of $\left(\frac{2(1+\varepsilon)}{1-\varepsilon}\right)$.
\end{Remark}
\subsection{Wringing Technique} \label{sectionWringing}
The following lemma forms part of the wringing technique proposed by Ahlswede and its proof can be found in \cite[Lemma 4]{Ahl82}.
\smallskip
\begin{Lemma}\label{lemmaFromAhlswede}
Let $\mathcal{X}$ be a finite alphabet, let $p_{X^n}$ and $u_{X^n}$ be two probability mass functions defined on $\mathcal{X}^n$ and let $c>0$ be a real number such that
\begin{equation}
p_{X^n}(x^n)\le (1+c)u_{X^n}(x^n)
\end{equation}
for all $x^n\in \mathcal{X}^n$. Fix any $0<\lambda <1$. Then for any $0<\delta<c$, there exist $\ell$ natural numbers in $\{1, 2, \ldots, n\}$, denoted by $t_1, t_2, \ldots, t_\ell$, and $\ell$ elements of $\mathcal{X}$ denoted by $\bar x_{t_1}, \bar x_{t_2}, \ldots, \bar x_{t_\ell}$, such that the following three statements hold:
 \begin{enumerate}
 \item[(I)] $\ell\le \frac{c}{\delta}$.
\item[(II)] $ \Pr_{p_{X^n}}\left\{(X_{t_1}, X_{t_2}, \ldots, X_{t_\ell})=(\bar x_{t_1}, \bar x_{t_2}, \ldots, \bar x_{t_\ell})\right\} \ge \lambda^\ell.$
 \item[(III)] For all $k\in\{1, 2, \ldots, n\}\setminus \{t_1, t_2, \ldots, t_\ell\}$, we have
 \begin{align}
&p_{X_k|X_{t_1}, X_{t_2}, \ldots, X_{t_\ell}}(x_k|\bar x_{t_1}, \bar x_{t_2}, \ldots, \bar x_{t_\ell}) \notag\\*
&\quad \le \,\max\{(1+\delta)u_{X_k|X_{t_1}, X_{t_2}, \ldots, X_{t_\ell}}(x_k|\bar x_{t_1}, \bar x_{t_2}, \ldots, \bar x_{t_\ell}), \lambda\} \label{lemmaFromAhlswedeSt1}
 \end{align}
 for all $x_k\in \mathcal{X}$.
 \end{enumerate}
\end{Lemma}
\smallskip

The crux of Lemma~\ref{lemmaFromAhlswede} is in the identification of the event \begin{equation}
\mathcal{F}\triangleq \{(X_{t_1}, X_{t_2}, \ldots, X_{t_\ell})=(\bar x_{t_1}, \bar x_{t_2}, \ldots, \bar x_{t_\ell})\}
  \end{equation}
  such that conditioned on~$\mathcal{F}$, the distributions of the resultant codeword symbols transmitted in each time slot~$k$ can be approximated by~$u_{X_k}$ (cf.~\eqref{lemmaFromAhlswedeSt1}). In the sequel where each $X_k$ in Lemma~\ref{lemmaFromAhlswede} is substituted by $\hat X_{T,k}$ where $\hat X_{T,k}$ is some quantized version of $X_{T,k}$ to be specified later, the joint distribution $u_{\hat X_{T,k}}$ that approximates $p_{\hat X_{T,k}}$ will be chosen to be a product distribution (cf.~\eqref{CorollaryWringingSt4}) with marginals $u_{\hat X_{i,k}}$.
In order to use Lemma~\ref{lemmaFromAhlswede} for proving Theorem~\ref{thmMainResult}, an important step involves controlling the size of $\mathcal{X}$ in Lemma~\ref{lemmaFromAhlswede}. To this end, we use the following scalar quantizer to quantize the alphabet $\mathcal{X}_i$ (in~\eqref{defAlphabetXi*}) which is exponential in the blocklength $n$ (cf.~\eqref{defAlphabetXi}) so that its quantized version is an alphabet  whose size is   polynomial in the blocklength.
\smallskip
\begin{Definition}\label{scalarQuantizer}
Let $L$ be a natural number and $\Delta$ be a positive real number, and let
\begin{equation}
\mathbb{Z}_{L,\Delta}\triangleq \{-L\Delta, (-L+1)\Delta, \ldots, L\Delta\} \label{defQuantizationSet}
 \end{equation}
 be a set of $2L+1$ quantization points where $\Delta$ specifies the quantization precision. A scalar quantizer with domain $[-L\Delta, L\Delta]$ and precision $\Delta$ is the mapping
\begin{equation}
\Omega_{L, \Delta}: [-L\Delta, L\Delta]\rightarrow \mathbb{Z}_{L,\Delta}
\end{equation}
such that
\begin{equation}
\Omega_{L, \Delta}(x) = \begin{cases}
\lfloor x/\Delta\rfloor \Delta & \text{ if $x\ge 0$,} \\
\lceil x/\Delta\rceil \Delta &\text{otherwise.}
\end{cases} \label{defOmega*}
\end{equation}
In other words, $\Omega_{L, \Delta}(x)$ maps $x$ to the closest quantized point whose value is smaller than or equal to $x$ if $x\ge 0$, and to the closest quantized point whose value is larger than or equal to $x$ if $x<0$. In addition, define the scalar quantizer for a real-valued tuple as
\begin{equation}
\Omega_{L, \Delta}^{(n)}: [-L\Delta, L\Delta]^n\rightarrow \mathbb{Z}_{L,\Delta}^n
\end{equation}
such that
\begin{equation}
\Omega_{L, \Delta}^{(n)}(x^n) \triangleq (\Omega_{L, \Delta}(x_1),\Omega_{L, \Delta}(x_2), \ldots, \Omega_{L, \Delta}(x_n)).
\end{equation}
\hfill $\blacksquare$
\end{Definition}
\smallskip

By our careful choice of the quantizer in Definition~\ref{scalarQuantizer}, we have the following property for all $x\in \mathbb{R}$:
\begin{align}
\left|\Omega_{L, \Delta}(x)\right| & \stackrel{\eqref{defOmega*}}{=}
\begin{cases}
\lfloor x/\Delta\rfloor \Delta & \text{ if $x\ge 0$,} \\
-\lceil x/\Delta\rceil \Delta &\text{otherwise}
\end{cases}
\\
& = \begin{cases}
\lfloor x/\Delta\rfloor \Delta & \text{ if $x\ge 0$,} \\
\lfloor -x/\Delta\rfloor \Delta &\text{otherwise}
\end{cases} \\
& = \lfloor |x|/\Delta\rfloor \Delta \\
& \le |x|.
\label{defOmega}
\end{align}

Although the following lemma looks similar to \cite[Corollary 2]{Ahl82} and they both rely on Lemma~\ref{lemmaFromAhlswede}, the proof of the following lemma is more involved due to the additional consideration of the quantizer's precision and the quantized input symbols.
If the quantizer's precision is too small or too large, then the resultant bound obtained from the following lemma will not be useful in proving the strong converse. See Section~\ref{sectionDiscussionQuantizer} for a detailed discussion on the appropriate choice for the quantizer's precision.
\smallskip
\begin{Lemma} \label{lemmaWringing}
Suppose we are given an $\big(n, M_{\mathcal{I}}^{(n)}, P_\mathcal{I}, \mathcal{A}^\prime, T , \frac{1+\varepsilon}{2}\big)_{\text{max}}$-code such that
\begin{equation}
|\mathcal{A}_T^\prime| = |\mathcal{A}^\prime| \ge \left(\frac{1-\varepsilon}{2(1+\varepsilon)}\right)\prod_{i\in T}M_i^{(n)} \label{CorollaryWringingSt}
\end{equation}
and
\begin{equation}
p_{W_T}^\prime(w_T) \le \frac{1}{\prod_{i\in T}M_i^{(n)}}\cdot\left(\frac{2(1+\varepsilon)}{1-\varepsilon}\right)
 \label{st2CorollaryWringing}
\end{equation}
for each $w_T\in \mathcal{A}_T^\prime$ where $p_{W_\mathcal{I}, X_\mathcal{I}^n, Y^n, \hat W_\mathcal{I}}^\prime$
  denotes the probability distribution induced on the Gaussian MAC by the $\big(n, M_{\mathcal{I}}^{(n)}, P_\mathcal{I}, \mathcal{A}^\prime, T, \frac{1+\varepsilon}{2}\big)_{\text{max}}$-code. Then, there exists an $\big(n, M_{\mathcal{I}}^{(n)}, P_\mathcal{I}, \mathcal{A}, T , \frac{1+\varepsilon}{2}\big)_{\text{max}}$-code with
\begin{equation}
|\mathcal{A}_T| = |\mathcal{A}| \ge n^{\frac{-4|T|(1+3\varepsilon)}{(1-\varepsilon)} \sqrt{ \frac{n}{\log n}} }\left(\frac{1-\varepsilon}{2(1+\varepsilon)}\right)\prod_{i\in T}M_i^{(n)} \label{CorollaryWringingSt3}
\end{equation}
such that the following holds: Let $p_{W_\mathcal{I}, X_\mathcal{I}^n, Y^n, \hat W_\mathcal{I}}$
  denote the probability distribution induced on the Gaussian MAC by the $\big(n, M_{\mathcal{I}}^{(n)}, P_\mathcal{I}, \mathcal{A}, T, \frac{1+\varepsilon}{2}\big)_{\text{max}}$-code. In addition, let
 \begin{equation}
 \hat X_i^n = \Omega_{\left\lceil n\sqrt{n P_i} \right\rceil,n^{-1}}^{(n)}(X_i^n), \label{defHatXin}
 \end{equation}
define the alphabet
 \begin{equation}
 \hat{\mathcal{X}}_i \triangleq \mathbb{Z}_{\left\lceil n\sqrt{n P_i} \right\rceil,n^{-1}} \label{defAlphabetXiInWringingLemma}
 \end{equation}
   for each $i \in T$ ($\hat X_i^n$ is always in the domain of $\mathbb{Z}_{\left\lceil n\sqrt{n P_i} \right\rceil,n^{-1}}^n$ because of \eqref{defHatXin}, \eqref{defOmega} and \eqref{powerConstraint}, and hence $\hat X_i^n \in \hat{\mathcal{X}}_i^n$),
   define
   \begin{equation}
   \hat{\mathcal{X}}_T \triangleq \prod_{i\in T}\hat{\mathcal{X}}_i
   \end{equation}
    and define
 \begin{align}
 & p_{W_\mathcal{I}, X_\mathcal{I}^n, \hat X_T^n, Y^n, \hat W_\mathcal{I}}(w_\mathcal{I}, x_\mathcal{I}^n, \hat x_T^n, y^n, \hat w_\mathcal{I}) \notag\\*
 &\quad \triangleq p_{W_\mathcal{I}, X_\mathcal{I}^n, Y^n, \hat W_\mathcal{I}}(w_\mathcal{I}, x_\mathcal{I}^n, y^n, \hat w_\mathcal{I}) \prod_{i\in T} \mathbf{1}\left\{\hat x_i^n = \Omega_{\left\lceil n\sqrt{n P_i} \right\rceil,n^{-1}}^{(n)}(x_i^n) \right\} \label{defQuantizedDistribution}
 \end{align}
for all
 $(w_\mathcal{I}, x_\mathcal{I}^n, \hat x_T^n, y^n, \hat w_\mathcal{I}) \in \mathcal{A}\times \mathcal{X}_\mathcal{I}^n \times \hat{\mathcal{X}}_T^n \times \mathbb{R}^n \times \mathcal{A}$. 
 Then there exists a distribution $u_{\hat X_T^n}$ defined on $\hat{\mathcal{X}}_T^n$ where
 \begin{align}
|\hat{\mathcal{X}}_T|\le n^{\frac{3|T|}{2}}\prod_{i\in T}(2\sqrt{P_i}+3) \label{CorollaryWringingSt7}
 \end{align}
  such that for all $k\in\{1, 2, \ldots, n\}$, we have
 \begin{align}
p_{\hat X_{T, k}}(\hat x_{T,k}) \le \max\left\{\left(1+\sqrt{\frac{\log n}{n}}\right)\prod_{i\in T}u_{\hat X_{i,k}}(\hat x_{i,k}), \frac{1}{n^{4|T|}}\right\} \label{CorollaryWringingSt4}
 \end{align}
 for all $\hat x_{T,k}\in \hat{\mathcal{X}}_T$ and
 \begin{equation}
 \sum_{i\in T}\sum_{k=1}^n \E_{u_{\hat X_{i,k}}}\left[\hat X_{i,k}^2\right] \le \sum_{i\in T}nP_i. \label{CorollaryWringingSt5}
 \end{equation}
\end{Lemma}

Before presenting the proof of Lemma~\ref{lemmaWringing}, we would like to stress the following two important implications of Lemma~\ref{lemmaWringing}.
\begin{enumerate}
\item[(i)]By identifying a certain event
\begin{equation}
\mathcal{G}\triangleq\{(\hat X_{T, t_1}, \hat X_{T, t_2}, \ldots, \hat X_{T, t_\ell})=(\bar x_{T, t_1}, \bar x_{T, t_2}, \ldots, \bar x_{T, t_\ell})\}
\end{equation}
(whose probability is quantified in~\eqref{eqn:statement(II)} in the following proof), we can find a subcode such that for each time slot~$k$, the resultant probability distribution of the quantized vector of transmitted symbols~$\hat X_{T,k}=(\hat X_{i,k}\,|\,i\in T)$ can be approximated by a product distribution $\prod_{i\in T}u_{\hat X_{i,k}}$ as in~\eqref{CorollaryWringingSt4}. This is the essence of the wringing technique~\cite{dueck81,Ahl82} which involves approximating the joint distribution of the random variables corresponding to the different encoders with a product distribution. By approximating $\hat X_{T,k}$ with a product distribution, we effectively \emph{wring} out the dependence among the collection of random variables $\{\hat X_{i,k}\,|\, i\in T\}$.
     \item[(ii)] The alphabet size of the quantized transmitted symbol $\hat X_{T,k}$ grows no faster than polynomially in~$n$ as in~\eqref{CorollaryWringingSt7}. Our quantization strategy that results in the polynomial growth of the alphabet sizes of the quantized symbols appears to be an important and necessary step, because the original alphabet size $|\mathcal{X}_T|$ could be exponentially large in~$n$ (cf.~\eqref{defAlphabetXi}). Furthermore, the controlled growth of $|\hat{\mathcal{X}}_T|$ ensures that $\Pr\{\mathcal{G}\}$ does not decay to zero exponentially fast as shown in~\eqref{eqn:statement(II)} in the following proof and hence the asymptotic rates of the resultant subcode are the same as that of the original code. An important point to note here is the following: We are able to lower bound the probability $\Pr\{\mathcal{G}\}$   because we defined $\mathcal{G}$ in terms of the \emph{quantized} random variables (rather than the original ones). The application of the wringing technique on the quantized random variables is one of the major contributions of the present work.
         \end{enumerate}
         \smallskip
\begin{IEEEproof}[Proof of Lemma~\ref{lemmaWringing}]
Let $p_{W_\mathcal{I}, X_\mathcal{I}^n, Y^n, \hat W_\mathcal{I}}^\prime$ be the probability distribution induced on the Gaussian MAC by the $\big(n, M_{\mathcal{I}}^{(n)},  P_\mathcal{I}, \mathcal{A}^\prime, T, \frac{1+\varepsilon}{2}\big)_{\text{max}}$-code that satisfies \eqref{CorollaryWringingSt} and \eqref{st2CorollaryWringing}, and let
\begin{align}
 & p_{W_\mathcal{I}, X_\mathcal{I}^n, \hat X_T^n, Y^n, \hat W_\mathcal{I}}^\prime(w_\mathcal{I}, x_\mathcal{I}^n, \hat x_T^n, y^n, \hat w_\mathcal{I}) \notag\\
 &\quad \triangleq p_{W_\mathcal{I}, X_\mathcal{I}^n, Y^n, \hat W_\mathcal{I}}^\prime(w_\mathcal{I}, x_\mathcal{I}^n, y^n, \hat w_\mathcal{I}) \prod_{i\in T} \mathbf{1}\left\{\hat x_i^n = \Omega_{\left\lceil n\sqrt{n P_i} \right\rceil,n^{-1}}^{(n)}(x_i^n) \right\}. \label{defQuantizedDistributionPprime}
 \end{align}
Define a probability mass function $u_{W_T, X_T^n, \hat X_T^n}^\prime$ as
\begin{equation}
u_{W_T, X_T^n, \hat X_T^n}^\prime(w_T, x_T^n, \hat x_T^n)
  \triangleq \prod_{i\in T} \frac{\mathbf{1}\left\{x_i^n = f_i(w_i) \right\} \cdot \mathbf{1}\left\{\hat x_i^n = \Omega_{\left\lceil n\sqrt{n P_i} \right\rceil,n^{-1}}^{(n)}(x_i^n) \right\}}{ M_i^{(n)}}\label{defDistUprime}
\end{equation}
 for all $(w_T, x_T^n, \hat x_T^n)\in \mathcal{W}_T\times \mathcal{X}_T^{n} \times \hat{\mathcal{X}}_T^n$ (cf.\ \eqref{defAlphabetXi*} and \eqref{defAlphabetXiInWringingLemma}), where $f_{i}$ represents the encoding function for $W_i$ of the $\big(n, M_{\mathcal{I}}^{(n)}, P_\mathcal{I}, \mathcal{A}^\prime, T, \frac{1+\varepsilon}{2}\big)_{\text{max}}$-code (cf.\ Definition~\ref{defCode}). The distribution $u_{W_T, X_T^n, \hat X_T^n}^\prime$ is well-defined (the probability masses sum to one) through \eqref{defDistUprime} because
 \begin{align}
& \sum\limits_{\substack{(w_T, x_T^n, \hat x_T^n)\in \\ \mathcal{W}_T\times \mathcal{X}_T^{n} \times \hat{\mathcal{X}}_T^n}} u_{W_T, X_T^n, \hat X_T^n}^\prime(w_T, x_T^n, \hat x_T^n) \\*
 & \quad \stackrel{\eqref{defDistUprime}}{=} \sum\limits_{\substack{w_T\in \mathcal{W}_T}}\prod_{i\in T} \frac{1}{M_i^{(n)}} \sum\limits_{\substack{x_T^n\in \mathcal{X}_T^{n}}}\prod_{i\in T} \mathbf{1}\left\{x_i^n = f_i(w_i) \right\} \sum\limits_{\hat x_T^n\in \hat{\mathcal{X}}_T^n}\prod_{i\in T}\mathbf{1}\left\{\hat x_i^n = \Omega_{\left\lceil n\sqrt{n P_i} \right\rceil,n^{-1}}^{(n)}(x_i^n) \right\} \\
 & \quad =1.
 \end{align}
Using \eqref{defDistUprime}, we obtain
 \begin{equation}
u_{W_T, X_T^n, \hat X_T^n}^\prime = \prod_{i\in T} u_{W_i, X_i^n, \hat X_i^n}^\prime \label{distUproductForm}
\end{equation}
where
\begin{equation}
u_{W_i, X_i^n, \hat X_i^n}^\prime(w_i, x_i^n, \hat x_i^n) = \frac{1}{M_i^{(n)}}\cdot \mathbf{1}\left\{x_i^n = f_i(w_i) \right\} \cdot \mathbf{1}\left\{\hat x_i^n = \Omega_{\left\lceil n\sqrt{n P_i} \right\rceil,n^{-1}}^{(n)}(x_i^n) \right\}
\end{equation}
for all $(w_i, x_i^n, \hat x_i^n)\in \mathcal{W}_i\times \mathcal{X}_i^{n} \times \hat{\mathcal{X}}_i^n$.
We will use Lemma~\ref{lemmaFromAhlswede} to prove the existence of a subcode of the $\big(n, M_{\mathcal{I}}^{(n)}, P_\mathcal{I}, \mathcal{A}^\prime, T, \frac{1+\varepsilon}{2}\big)_{\text{max}}$-code such that the subcode satisfies \eqref{CorollaryWringingSt3}, \eqref{CorollaryWringingSt4} and \eqref{CorollaryWringingSt5} for some $u_{\hat X_{T}^n}$ defined on~$\hat{\mathcal{X}}_T^n$. To this end, we first consider the following chain of inequalities for each $\hat x_T^n \in \hat{\mathcal{X}}_T^n$ such that $ p_{\hat X_T^n}^\prime(\hat x_T^n)>0$:
 \begin{align}
 p_{\hat X_T^n}^\prime(\hat x_T^n) & = \sum_{w_T\in \mathcal{A}_T^\prime, x_T^n\in \mathcal{X}_T^n} p_{W_T, X_T^n, \hat X_T^n}^\prime(w_T, x_T^n,\hat x_T^n) \\
 & = \sum_{w_T\in \mathcal{A}_T^\prime, x_T^n\in \mathcal{X}_T^n} p_{W_T}^\prime(w_T)p_{X_T^n, \hat X_T^n|W_T}^\prime(x_T^n,\hat x_T^n|w_T) \\
 &\stackrel{\text{(a)}}{=} \sum_{w_T\in \mathcal{A}_T^\prime, x_T^n\in \mathcal{X}_T^n} p_{W_T}^\prime(w_T) \prod_{i\in T}\left(\mathbf{1}\left\{x_i^n = f_i(w_i) \right\} \cdot \mathbf{1}\left\{\hat x_i^n = \Omega_{\left\lceil n\sqrt{n P_i} \right\rceil,n^{-1}}^{(n)}(x_i^n) \right\}\right) \\
 &\stackrel{\eqref{st2CorollaryWringing}}{\le} \sum_{w_T\in \mathcal{A}_T^\prime, x_T^n\in \mathcal{X}_T^n}\frac{1}{\prod_{i\in T}M_i^{(n)}}\cdot\left(\frac{2(1+\varepsilon)}{1-\varepsilon}\right)\prod_{i\in T}\left(\mathbf{1}\left\{x_i^n = f_i(w_i) \right\} \cdot \mathbf{1}\left\{\hat x_i^n = \Omega_{\left\lceil n\sqrt{n P_i} \right\rceil,n^{-1}}^{(n)}(x_i^n) \right\}\right) \\
 & \stackrel{\eqref{defDistUprime}}{=} \frac{2(1+\varepsilon)}{1-\varepsilon}\sum_{w_T\in \mathcal{A}_T^\prime, x_T^n\in \mathcal{X}_T^n}u_{W_T, X_T^n, \hat X_T^n}^\prime(w_T, x_T^n, \hat x_T^n) \\
 & \le \frac{2(1+\varepsilon)}{1-\varepsilon}\sum_{w_T\in \mathcal{W}_T, x_T^n\in \mathcal{X}_T^n}u_{W_T, X_T^n, \hat X_T^n}^\prime(w_T, x_T^n, \hat x_T^n) \\
 & = \frac{2(1+\varepsilon)}{1-\varepsilon}\cdot u_{\hat X_T^n}^\prime(\hat x_T^n) \label{corollaryProofSt1}
 \end{align}
 where (a) follows from \eqref{memorylessStatement*} and \eqref{defQuantizedDistributionPprime}.
 It follows from \eqref{corollaryProofSt1} and Lemma~\ref{lemmaFromAhlswede} with the identifications
\begin{equation}
 \mathcal{X}\triangleq \hat{\mathcal{X}}_T,\qquad c\triangleq \frac{1+3\varepsilon}{1-\varepsilon},\qquad \lambda\triangleq \frac{1}{n^{4|T|}},\qquad \delta\triangleq \sqrt{\frac{\log n}{n}} \label{identification1}
 \end{equation}
that there exist $\ell$ natural numbers in $\{1, 2, \ldots, n\}$, denoted by $t_1, t_2, \ldots, t_\ell$, and $\ell$ real-valued $|T|$-dimensional tuples in $\hat{\mathcal{X}}_T$, denoted by $\bar x_{T, t_1}, \bar x_{T, t_2}, \ldots, \bar x_{T, t_\ell}$, such that the following three statements hold:
 \begin{enumerate}
 \item[(I)] $\ell\le \left(\frac{1+3\varepsilon}{1-\varepsilon}\right)\sqrt{\frac{n}{\log n}}$.
 \item[(II)] \hfill \makebox[0pt][r]{%
            \begin{minipage}[b]{1.29\textwidth}
              \begin{equation}
              \Pr_{p_{\hat X^n}^\prime}\left\{(\hat X_{T, t_1}, \hat X_{T, t_2}, \ldots, \hat X_{T, t_\ell})=(\bar x_{T, t_1}, \bar x_{T, t_2}, \ldots, \bar x_{T, t_\ell})\right\} \ge \frac{1}{n^{4|T|\ell}} . \label{eqn:statement(II)}
              \end{equation}
          \end{minipage}}
 \item[(III)] For all $k\in\{1, 2, \ldots, n\}\setminus \{t_1, t_2, \ldots, t_\ell\}$, we have
 \begin{align}
&p_{\hat X_{T,k}|\hat X_{T, t_1}, \hat X_{T, t_2}, \ldots, \hat X_{T, t_\ell}}^\prime(\hat x_{T,k}|\bar x_{T, t_1}, \bar x_{T, t_2}, \ldots, \bar x_{T, t_\ell}) \notag\\*
&\quad \le \,\max\left\{\left(1+\sqrt{\frac{\log n}{n}}\right)u_{\hat X_{T,k}|\hat X_{T, t_1}, \hat X_{T, t_2}, \ldots, \hat X_{T, t_\ell}}^\prime(\hat x_{T,k}|\bar x_{T, t_1}, \bar x_{T, t_2}, \ldots, \bar x_{T, t_\ell}), \frac{1}{n^{4|T|}}\right\} \\
& \quad \stackrel{\eqref{distUproductForm}}{=} \,\max\left\{\left(1+\sqrt{\frac{\log n}{n}}\right)\prod_{i\in T}u_{\hat X_{i,k}|\hat X_{i, t_1}, \hat X_{i, t_2}, \ldots, \hat X_{i, t_\ell}}^\prime(\hat x_{i,k}|\bar x_{i, t_1}, \bar x_{i, t_2}, \ldots, \bar x_{i, t_\ell}), \frac{1}{n^{4|T|}}\right\}
 \end{align}
 for all $\hat x_{T,k}\in \hat{\mathcal{X}}_T$.
 \end{enumerate}
Using Statement (II), Statement (III) and \eqref{CorollaryWringingSt}, we can construct an $\big(n, M_{\mathcal{I}}^{(n)}, P_\mathcal{I}, \mathcal{A}, T , \frac{1+\varepsilon}{2}\big)_{\text{max}}$-code by collecting all the codewords $x_\mathcal{I}^n$ for the $\big(n, M_{\mathcal{I}}^{(n)}, P_\mathcal{I}, \mathcal{A}^\prime, T , \frac{1+\varepsilon}{2}\big)_{\text{max}}$-code which satisfy
\begin{equation}
(\hat x_{T, t_1}, \hat x_{T, t_2}, \ldots, \hat x_{T, t_\ell})= (\bar x_{T, t_1}, \bar x_{T, t_2}, \ldots, \bar x_{T, t_\ell})
\end{equation}
 such that the following two statements hold:
\begin{enumerate}
\item[(i)] $|\mathcal{A}_T| = |\mathcal{A}| \ge n^{-4|T|\ell}\left(\frac{1-\varepsilon}{2(1+\varepsilon)}\right)\prod_{i\in T}M_i^{(n)}$.
\item[(ii)] Let $p_{W_\mathcal{I}, X_\mathcal{I}^n, Y^n, \hat W_\mathcal{I}}$
  denote the probability distribution induced on the Gaussian MAC by the $\big(n, M_{\mathcal{I}}^{(n)}, P_\mathcal{I}, \mathcal{A}, T, \linebreak \frac{1+\varepsilon}{2}\big)_{\text{max}}$-code, and let
 \begin{align}
 & p_{W_\mathcal{I}, X_\mathcal{I}^n, \hat X_T^n, Y^n, \hat W_\mathcal{I}}(w_\mathcal{I}, x_\mathcal{I}^n, \hat x_T^n, y^n, \hat w_\mathcal{I}) \notag\\
 &\quad \triangleq p_{W_\mathcal{I}, X_\mathcal{I}^n, Y^n, \hat W_\mathcal{I}}(w_\mathcal{I}, x_\mathcal{I}^n, y^n, \hat w_\mathcal{I}) \prod_{i\in T} \mathbf{1}\left\{\hat x_i^n = \Omega_{\left\lceil n\sqrt{n P_i} \right\rceil,n^{-1}}^{(n)}(x_i^n) \right\}. \label{defQuantizedDistributionInProof}
 \end{align}
 Then,
 \begin{equation}
 \Pr_{p_{\hat X_T^n}}\left\{\bigcap_{m=1}^\ell \{\hat X_{T,t_m} = \bar x_{T, t_m}\} \right\}=1, \label{eqn:statement(ii)inProof}
 \end{equation}
 and we have for all $k\in\{1, 2, \ldots, n\} \setminus \{t_1, t_2, \ldots, t_\ell\}$
\begin{align}
p_{\hat X_{T,k}}(\hat x_{T,k}) \le \,\max\left\{\left(1+\sqrt{\frac{\log n}{n}}\right)\prod_{i\in T}u_{\hat X_{i,k}|\hat X_{i, t_1}=\bar x_{i, t_1}, \hat X_{i, t_2}=\bar x_{i, t_2}, \ldots, \hat X_{i, t_\ell}=\bar x_{i, t_\ell}}^\prime(\hat x_{i,k}), \frac{1}{n^{4|T|}}\right\} \label{eqn:statement(ii)inProof*}
\end{align}
 for all $\hat x_{T,k}\in \hat{\mathcal{X}}_T$.
\end{enumerate}
Since for each $k\in\{t_1, t_2, \ldots, t_\ell\}$
\begin{align}
p_{\hat X_{T,k}}(\hat x_{T,k}) \stackrel{\eqref{eqn:statement(ii)inProof}}{=} \mathbf{1}\left\{ \hat x_{T,k} = \bar x_{T, k} \right\} = \prod_{i\in T}u_{\hat X_{i,k}|\hat X_{i, t_1}=\bar x_{i, t_1}, \hat X_{i, t_2}=\bar x_{i, t_2}, \ldots, \hat X_{i, t_\ell}=\bar x_{i, t_\ell}}^\prime(\hat x_{i,k})
\end{align}
 for all $\hat x_{T,k}\in \hat{\mathcal{X}}_T$, it follows from \eqref{eqn:statement(ii)inProof*} that the following statement holds:
 \begin{enumerate}
 \item[(iii)] For all $k\in\{1, 2, \ldots, n\}$, we have
\begin{equation}
 p_{\hat X_{T,k}}(\hat x_{T,k}) \le \,\max\left\{\left(1+\sqrt{\frac{\log n}{n}}\right)\prod_{i\in T}u_{\hat X_{i,k}|\hat X_{i, t_1}=\bar x_{i, t_1}, \hat X_{i, t_2}=\bar x_{i, t_2}, \ldots, \hat X_{i, t_\ell}=\bar x_{i, t_\ell}}^\prime(\hat x_{i,k}), \frac{1}{n^{4|T|}}\right\}
 \end{equation}
for all $\hat x_{T,k}\in \hat{\mathcal{X}}_T$.
 \end{enumerate}
Consequently, \eqref{CorollaryWringingSt3} follows from Statement~(i) and Statement~(I), and \eqref{CorollaryWringingSt4} follows from Statement~(iii) by letting
 \begin{equation}
 u_{\hat X_{T}^n} \triangleq \prod_{k=1}^n \prod_{i\in T}u_{\hat X_{i,k}|\hat X_{i, t_1}=\bar x_{i, t_1}, \hat X_{i, t_2}=\bar x_{i, t_2}, \ldots, \hat X_{i, t_\ell}=\bar x_{i, t_\ell}}^\prime. \label{defU}
 \end{equation}

It remains to prove the upper bounds on $|\hat{\mathcal{X}}_T|$ and $\sum_{i\in T}\sum_{k=1}^n\E_{u_{\hat X_{i,k}}}\left[\hat X_{i,k}^2\right]$ in~\eqref{CorollaryWringingSt7} and~\eqref{CorollaryWringingSt5} respectively.
To prove~\eqref{CorollaryWringingSt7}, we consider
\begin{align}
|\hat{\mathcal{X}}_T| & \stackrel{\eqref{defAlphabetXiInWringingLemma}}{=} \prod_{i\in T} \left(2\left\lceil n\sqrt{n P_i} \right\rceil + 1\right) \\
& \le  \prod_{i\in T} \left(2n^{3/2}\sqrt{P_i} + 3\right)\\
& \le n^{\frac{3|T|}{2}}\prod_{i\in T} (2\sqrt{P_i}+3).
\end{align}
To prove~\eqref{CorollaryWringingSt5}, we first use \eqref{defDistUprime} and \eqref{powerConstraint} to obtain
 \begin{equation}
 \Pr_{u_{X_T^n, \hat X_T^n}^\prime}\left\{\sum_{i\in T}\sum_{k=1}^n X_{i,k}^2 \le \sum_{i\in T}nP_i \right\}=1. \label{uPrimeXPowerConstraint}
 \end{equation}
 Since $\hat X_{i,k}^2 \le X_{i,k}^2$ for all $i \in T$ and all $k\in\{1, 2, \ldots, n\}$ by \eqref{defQuantizedDistributionPprime} and \eqref{defOmega}, it follows from \eqref{uPrimeXPowerConstraint} that
 \begin{equation}
 \Pr_{u_{\hat X_T^n}^\prime}\left\{\sum_{i\in T}\sum_{k=1}^n \hat X_{i,k}^2 \le \sum_{i\in T}nP_i \right\}=1. \label{uPrimeXHatPowerConstraint}
 \end{equation}
Consequently,
 \begin{align}
 & \sum_{i\in T}\sum_{k=1}^n\E_{u_{\hat X_{i,k}}}\left[\hat X_{i,k}^2\right]\notag\\
 &\quad \stackrel{\eqref{defU}}{=} \sum_{i\in T}\sum_{k=1}^n\E_{u_{\hat X_{i,k}|\hat X_{i, t_1}=\bar x_{i, t_1}, \hat X_{i, t_2}=\bar x_{i, t_2}, \ldots, \hat X_{i, t_\ell}=\bar x_{i, t_\ell}}^\prime}\left[\hat X_{i,k}^2\right]\\
 & \quad \stackrel{\eqref{distUproductForm}}{=}\sum_{i\in T}\sum_{k=1}^n\E_{u_{\hat X_{T,k}|\hat X_{T, t_1}=\bar x_{T, t_1}, \hat X_{T, t_2}=\bar x_{T, t_2}, \ldots, \hat X_{T, t_\ell}=\bar x_{T, t_\ell}}^\prime}\left[\hat X_{i,k}^2\right]\\
 &\quad =\sum_{i\in T}\sum_{k=1}^n\E_{u_{\hat X_T^n|\hat X_{T, t_1}=\bar x_{T, t_1}, \hat X_{T, t_2}=\bar x_{T, t_2}, \ldots, \hat X_{T, t_\ell}=\bar x_{T, t_\ell}}^\prime}\left[\hat X_{i,k}^2\right]\\
 &\quad =\E_{u_{\hat X_T^n|\hat X_{T, t_1}=\bar x_{T, t_1}, \hat X_{T, t_2}=\bar x_{T, t_2}, \ldots, \hat X_{T, t_\ell}=\bar x_{T, t_\ell}}^\prime}\left[\sum_{i\in T}\sum_{k=1}^n\hat X_{i,k}^2\right]\\
 &\quad \stackrel{\eqref{uPrimeXHatPowerConstraint}}{\le}\sum_{i\in T} nP_i\,.
 \end{align}
\end{IEEEproof}

\subsection{Binary Hypothesis Testing} \label{sectionBHT}
The following definition concerning the non-asymptotic fundamental limits of a simple binary hypothesis test is standard. See for example \cite[Sec.~III-E]{PPV10}.
\medskip
\begin{Definition}\label{defBHTDivergence}
Let $p_{X}$ and $q_{X}$ be two probability distributions on some common alphabet $\mathcal{X}$. Let
\[
\mathcal{Q}(\{0,1\}|\mathcal{X})\triangleq \{
r_{Z|X} \,|\, \text{$Z$ and $X$ assume values in $\{0,1\}$ and $\mathcal{X}$ respectively}\}
\]
be the set of randomized binary hypothesis tests between $p_{X}$ and $q_{X}$ where $\{Z=0\}$ indicates the test chooses $q_X$, and let $\delta\in [0,1]$ be a real number. The minimum type-II error in a simple binary hypothesis test between $p_{X}$ and $q_{X}$ with type-I error no larger than $1-\delta$ is defined as
\begin{align}
 \beta_{\delta}(p_X\|q_X) \triangleq
\inf\limits_{\substack{r_{Z|X} \in \mathcal{Q}(\{0,1\}|\mathcal{X}): \\ \int_{x\in \mathcal{X}}r_{Z|X}(1|x)p_X(x)\, \mathrm{d}x\ge \delta}} \int_{x\in \mathcal{X}}r_{Z|X}(1|x)q_X(x)\, \mathrm{d}x.\label{eqDefISDivergence}
\end{align}
\end{Definition}\medskip
The existence of a minimizing test $r_{Z|X}$ is guaranteed by the Neyman-Pearson lemma.

We state in the following lemma and proposition some important properties of $\beta_{\delta}(p_X\|q_X)$, which are crucial for the proof of Theorem~\ref{thmMainResult}. The proof of the following lemma can be found in, for example, the paper by Wang, Colbeck, and Renner~\cite[Lemma~1]{wang09}.
\medskip
\begin{Lemma}\label{lemmaDPI} Let $p_{X}$ and $q_{X}$ be two probability distributions on some alphabet $\mathcal{X}$, and let $g$ be a function whose domain contains $\mathcal{X}$. Then, the following two statements hold:
\begin{enumerate}
\item[1.] Data processing inequality (DPI):
\begin{equation}
\beta_{\delta}(p_X\|q_X) \le \beta_{\delta}(p_{g(X)}\|q_{g(X)}).
\end{equation}

\item[2.] For all $\xi>0$,
\begin{equation}
\beta_{\delta}(p_X\|q_X)\ge \frac{1}{\xi}\left(\delta - \int_{x\in\mathcal{X}}p_X(x) \boldsymbol{1}\left\{ \frac{p_X(x)}{q_X(x)} \ge \xi \right\}\, \mathrm{d}x\right) .
\end{equation}
\end{enumerate}
\end{Lemma}
\medskip
 The proof of the following proposition is similar to Lemma~3 in \cite{wang09} and therefore omitted.
 \medskip
\begin{Proposition} \label{propositionBHTLowerBound}
Let $p_{U,V}$ be a probability distribution defined on $\mathcal{W}\times \mathcal{W}$ for some finite alphabet $\mathcal{W}$. In addition, let $q_{V}$ be a distribution defined on $\mathcal{W}$, and let
\begin{equation}
\alpha = \max_{u\in\mathcal{W}} \Pr\{V\ne u|U=u\} \label{defAlpha}
\end{equation}
be a real number in $[0, 1)$ where $(U,V)$ is distributed according to $p_{U,V}$. Then for each $u\in\mathcal{W}$,
\begin{equation}
\beta_{1-\alpha}(p_{V|U=u}\|q_{V}) \le q_{V}(u). \label{propositionBHTLowerBoundEq1}
\end{equation}
\end{Proposition}

\section{Proof of Theorem~\ref{thmMainResult}} \label{sec:prf_main_result}
\subsection{Expurgation to Obtain a Maximum Error Code}
 Let $\varepsilon\in[0,1)$ and suppose $R_\mathcal{I}$ is an $\varepsilon$-achievable rate tuple. By Definition~\ref{defAchievableRate}, there exists a $\gamma \in [0,1)$ and a sequence of $(n, M_{\mathcal{I}}^{(n)}, P_\mathcal{I}, \varepsilon_n)_{\text{avg}}$-codes such that
 \begin{equation}
 \varepsilon_n \le \gamma \label{errorProbabilityInProof}
 \end{equation}
 for all sufficiently large~$n$ and
 \begin{equation}
\liminf\limits_{n\rightarrow \infty}\frac{1}{n}\log M_i^{(n)} \ge R_i \label{achievableRateInProof}
 \end{equation}
for each $i\in\mathcal{I}$. Fix a non-empty set $T\subseteq \mathcal{I}$. Our goal is to prove that
 \begin{align}
 \sum_{i\in T} R_i \le \frac{1}{2}\log\bigg(1+\sum_{i\in T}P_i\bigg) \label{goalInProof}.
 \end{align}
 Since \eqref{goalInProof} holds trivially if $\sum_{i\in T} R_i=0$, we assume without loss of generality that
 \begin{align}
 \sum_{i\in T} R_i >0. \label{assumption1InProof}
 \end{align}
 It follows from \eqref{achievableRateInProof} and \eqref{assumption1InProof} that
 \begin{align}
 \left\lfloor \left(\frac{1-\gamma}{1+\gamma}\right)\prod_{i\in T}M_i^{(n)} \right\rfloor \ge \frac{1}{2} \left(\frac{1-\gamma}{1+\gamma}\right)\prod_{i\in T}M_i^{(n)} \label{assumption3InProof}
 \end{align}
 for all sufficiently large~$n$. Fix a sufficiently large~$n$ and the corresponding $(n, M_{\mathcal{I}}^{(n)}, P_\mathcal{I}, \varepsilon_n)_{\text{avg}}$-code for the Gaussian MAC such that \eqref{errorProbabilityInProof} and \eqref{assumption3InProof} hold.
Using Lemma~\ref{lemmaExpurgation}, Lemma~\ref{lemmaWringing} and Definition~\ref{defCode}, there exists an $\big(n, M_\mathcal{I}^{(n)}, P_\mathcal{I}, \mathcal{A}, T, \frac{1+\gamma}{2}\big)_{\text{max}}$-code, which induces a probability distribution on the Gaussian MAC denoted by $p_{W_\mathcal{I}, X_\mathcal{I}^n, Y^n, \hat W_\mathcal{I}}$, such that the following four statements hold:
\begin{enumerate}
\item[(i)] For all $w_\mathcal{I}\in \mathcal{A}$ and all $w_T\in \mathcal{A}_T$,
\begin{equation}
p_{W_\mathcal{I}}(w_\mathcal{I}) = \frac{1}{|\mathcal{A}|}\text{ and }p_{W_T}(w_T) = \frac{1}{|\mathcal{A}_T|} . \label{eqn:statement(i)}
\end{equation}
\item[(ii)] There exists a $w_{T^c}^*\in \mathcal{W}_{T^c}$ such that for all $w_\mathcal{I}\in \mathcal{A}$, we have $w_{T^c}=w_{T^c}^*$.
\item[(iii)] The support of $W_T$ satisfies
 \begin{equation}
|\mathcal{A}_T| = |\mathcal{A}| \ge n^{\frac{-4|T|(1+3\gamma)}{(1-\gamma)} \sqrt{\frac{n}{\log n}} }\left(\frac{1-\gamma}{2(1+\gamma)}\right)\prod_{i\in T}M_i^{(n)}. \label{eqn:statement(ii)}
 \end{equation}
\item[(iv)] Define
 \begin{align}
 & p_{W_\mathcal{I}, X_\mathcal{I}^n, \hat X_T^n, Y^n, \hat W_\mathcal{I}}(w_\mathcal{I}, x_\mathcal{I}^n, \hat x_T^n, y^n, \hat w_\mathcal{I}) \notag\\
 &\quad \triangleq p_{W_\mathcal{I}, X_\mathcal{I}^n, Y^n, \hat W_\mathcal{I}}(w_\mathcal{I}, x_\mathcal{I}^n, y^n, \hat w_\mathcal{I}) \prod_{i\in T} \mathbf{1}\left\{\hat x_i^n = \Omega_{\left\lceil n\sqrt{n P_i} \right\rceil,n^{-1}}^{(n)}(x_i^n) \right\} \label{defQuantizedDistributionInMainProof}
 \end{align}
for all $(w_\mathcal{I}, x_\mathcal{I}^n, \hat x_T^n, y^n, \hat w_\mathcal{I})\in \mathcal{A}\times \mathcal{X}_\mathcal{I}^n \times \hat{\mathcal{X}}_T^n \times \mathbb{R}^n \times  \mathcal{A}$, where
\begin{equation}
\hat{\mathcal{X}}_T \triangleq \prod_{i\in T}\mathbb{Z}_{\left\lceil n\sqrt{n P_i} \right\rceil,n^{-1}} \label{eqn:quant_alpha}
\end{equation}
and
\begin{equation}
|\hat{\mathcal{X}}_T| \le n^{\frac{3|T|}{2}}\prod_{i\in T}(2\sqrt{P_i}+3).  \label{eqn:size_Xhat}
\end{equation}
 Then there exists a distribution $u_{\hat X_T^n}$ defined on $\hat{\mathcal{X}}_T^n$ such that for all $k\in\{1, 2, \ldots, n\}$, we have
 \begin{align}
p_{\hat X_{T, k}}(\hat x_{T,k}) \le \max\left\{\left(1+\sqrt{\frac{\log n}{n}}\right)\prod_{i\in T}u_{\hat X_{i,k}}(\hat x_{i,k}), \frac{1}{n^{4|T|}}\right\} \label{eqn:statement(iv)}
 \end{align}
 for all $\hat x_{T,k}\in \hat{\mathcal{X}}_T$ and
 \begin{align}
\sum_{i\in T} \sum_{k=1}^n \E_{u_{\hat X_{i}}} \left[\hat X_{i,k}^2\right] \le \sum_{i\in T}nP_i \, . \label{eqn:statement(v)}
 \end{align}
\end{enumerate}
Note that $p_{W_\mathcal{I}, X_\mathcal{I}^n, Y^n, \hat W_\mathcal{I}}$ is not the distribution induced by the original $(n, M_{\mathcal{I}}^{(n)}, P_\mathcal{I}, \varepsilon_n)_{\text{avg}}$-code but rather it is induced by the expurgated $\big(n, M_\mathcal{I}^{(n)}, P_\mathcal{I}, \mathcal{A}, T, \frac{1+\gamma}{2}\big)_{\text{max}}$-code.

\subsection{Lower Bounding the Error Probability using Binary Hypothesis Testing}
Now, let
\begin{equation}
s_{W_\mathcal{I}, X_\mathcal{I}^n, Y^n, \hat W_\mathcal{I}}\triangleq p_{W_\mathcal{I}, X_\mathcal{I}^n}\left(\prod_{k=1}^n s_{Y_k|X_{T^c,k}}\right) p_{\hat W_\mathcal{I}|Y^n} \label{defSimulatingDistSsumRate}
\end{equation}
be a distribution such that for each $k\in\{1, 2, \ldots, n\}$,  the auxiliary conditional output distribution is chosen to be
\begin{equation}
s_{Y_k|X_{T^c,k}}(y_k|x_{T^c,k}) = \mathcal{N}\left(y_k; \sum_{i\in T}\E_{u_{\hat X_{i,k}}}[\hat X_{i,k}] + \sum_{j\in T^c}x_{j,k},1+\sum_{i\in T}P_i \right) \label{defSimulatingDistSKsumRate}
\end{equation}
for all $x_{T^c,k} \in \mathcal{X}_{T^c}$ and $y_k\in\mathbb{R}$. It can be seen from \eqref{defSimulatingDistSsumRate} and \eqref{defSimulatingDistSKsumRate} that $s_{W_\mathcal{I}, X_\mathcal{I}^n, Y^n, \hat W_\mathcal{I}}$ depends on the choice of~$T$ we fixed at the start of the proof and the distribution $u_{\hat X_T^n}$ in Statement~(iv). We shall see later that this choice of $s_{W_\mathcal{I}, X_\mathcal{I}^n, Y^n, \hat W_\mathcal{I}}$, in particular the mean of the distribution in~\eqref{defSimulatingDistSKsumRate} namely $ \sum_{i\in T}\E_{u_{\hat X_{i,k}}}[\hat X_{i,k}] + \sum_{j\in T^c}x_{j,k}$, combined with Proposition~\ref{propositionBHTLowerBound} and Lemma~\ref{lemmaDPI} enables us to prove~\eqref{goalInProof}. We do not index $s_{W_\mathcal{I}, X_\mathcal{I}^n, Y^n, \hat W_\mathcal{I}}$ by $T$ nor $u_{\hat X_T^n}$ for notational brevity.
To simplify notation, let $\bar \gamma\triangleq (1+\gamma)/2$ be the maximal probability of decoding error of the $\big(n, M_\mathcal{I}^{(n)}, P_\mathcal{I}, \mathcal{A}, T, \frac{1+\gamma}{2}\big)_{\text{max}}$-code, where $\bar \gamma < 1$ because $\gamma<1$.
 Then for each $w_\mathcal{I} \in \mathcal{A}$, since
\begin{equation}
 s_{W_\mathcal{I}}(w_\mathcal{I})\stackrel{\eqref{defSimulatingDistSsumRate}}{=}p_{W_\mathcal{I}}(w_\mathcal{I})\stackrel{\eqref{eqn:statement(i)}}{>}0,
 \end{equation}
 it follows from Proposition~\ref{propositionBHTLowerBound} and Definition~\ref{defCode} with the identifications $U\equiv W_T$, $V\equiv \hat W_T$, $p_{U,V}\equiv p_{W_T, \hat W_T|W_{T^c}=w_{T^c}}$, $q_V\equiv s_{\hat W_T|W_{T^c}=w_{T^c}}$ and $\alpha\equiv \max_{w_\mathcal{I}\in \mathcal{A}} \Pr\{\hat W_T\ne w_T| W_\mathcal{I}= w_\mathcal{I}\} \le \bar\gamma$ that
 \begin{align}
&\beta_{1-\bar \gamma}(p_{\hat W_T|W_\mathcal{I}= w_\mathcal{I}}\|s_{\hat W_T|W_{T^c}=w_{T^c}}) \notag\\*
& \le \beta_{1-\alpha}(p_{\hat W_T|W_\mathcal{I}= w_\mathcal{I}}\|s_{\hat W_T|W_{T^c}=w_{T^c}})\\
&\le s_{\hat W_T|W_{T^c}}(w_T|w_{T^c}). \label{eqnBHTReverseChain}
 \end{align}
 \subsection{Using the DPI to Introduce the Channel Inputs and Output}
 Consider the following chain of inequalities for each $ w_\mathcal{I} \in \mathcal{A}$:
 \begin{align}
& \beta_{1-\bar \gamma}(p_{\hat W_\mathcal{I}|W_{\mathcal{I}}=w_{\mathcal{I}}}\|s_{\hat W_\mathcal{I}|W_{T^c}=w_{T^c}}) \notag \\
&\stackrel{\text{(a)}}{\ge} \beta_{1-\bar \gamma}(p_{Y^n,\hat W_\mathcal{I}|W_{\mathcal{I}}=w_{\mathcal{I}}}\|s_{Y^n,\hat W_\mathcal{I}|W_{T^c}=w_{T^c}}) \\
& = \beta_{1-\bar \gamma}(p_{Y^n|W_{\mathcal{I}}=w_{\mathcal{I}}}p_{\hat W_\mathcal{I}|Y^n, W_{\mathcal{I}}=w_{\mathcal{I}}}\|s_{Y^n,\hat W_\mathcal{I}|W_{T^c}=w_{T^c}}) \\
& \stackrel{\text{(b)}}{=} \beta_{1-\bar \gamma}(p_{Y^n|W_{\mathcal{I}}=w_{\mathcal{I}}}p_{\hat W_\mathcal{I}|Y^n}\|s_{Y^n,\hat W_\mathcal{I}|W_{T^c}=w_{T^c}}) \\
&\stackrel{\text{(c)}}{\ge} \beta_{1-\bar \gamma}\left(p_{\hat W_\mathcal{I}|Y^n}p_{X_\mathcal{I}^n, Y^n|W_{\mathcal{I}}=w_{\mathcal{I}}}\left\|p_{X_T^n | X_{T^c}^n, W_{\mathcal{I}}=w_{\mathcal{I}}}s_{X_{T^c}^n,Y^n,\hat W_\mathcal{I}|W_{T^c}=w_{T^c}}\right.\right) \\
& \stackrel{\eqref{defSimulatingDistSsumRate}}{=} \beta_{1-\bar \gamma}\left(p_{\hat W_\mathcal{I}|Y^n}p_{X_\mathcal{I}^n, Y^n|W_{\mathcal{I}}=w_{\mathcal{I}}}\left\|p_{X_T^n|X_{T^c}^n, W_{\mathcal{I}}=w_{\mathcal{I}}} p_{X_{T^c}^n|W_{T^c}=w_{T^c}} p_{\hat W_\mathcal{I}|Y^n}\prod_{k=1}^n s_{Y_k|X_{T^c,k}}\right.\right) \\
& \stackrel{\text{(d)}}{=}\beta_{1-\bar \gamma}\left(p_{\hat W_\mathcal{I}|Y^n}p_{X_\mathcal{I}^n, Y^n|W_{\mathcal{I}}=w_{\mathcal{I}}}\left\|p_{X_T^n|X_{T^c}^n, W_{\mathcal{I}}=w_{\mathcal{I}}} p_{X_{T^c}^n|W_{\mathcal{I}}=w_{\mathcal{I}}} p_{\hat W_\mathcal{I}|Y^n}\prod_{k=1}^n s_{Y_k|X_{T^c,k}}\right.\right) \\
& = \beta_{1-\bar \gamma}\left(p_{\hat W_\mathcal{I}|Y^n}p_{X_\mathcal{I}^n, Y^n|W_{\mathcal{I}}=w_{\mathcal{I}}}\left\|p_{X_\mathcal{I}^n|W_{\mathcal{I}}=w_{\mathcal{I}}} p_{\hat W_\mathcal{I}|Y^n}\prod_{k=1}^n s_{Y_k|X_{T^c,k}}\right.\right) \\
&\stackrel{\eqref{memorylessStatement*}}{=} \beta_{1-\bar \gamma}\left(p_{X_\mathcal{I}^n |W_{\mathcal{I}}=w_\mathcal{I}}p_{\hat W_\mathcal{I}|Y^n}\prod_{k=1}^n p_{Y_k|X_{\mathcal{I},k}}\left\|p_{X_\mathcal{I}^n |W_{\mathcal{I}}=w_\mathcal{I}}p_{\hat W_\mathcal{I}|Y^n}\prod_{k=1}^n s_{Y_k|X_{T^c,k}}\right.\right), \label{eqnBHTFirstChain}
 \end{align}
 where
 \begin{enumerate}
 \item[(a)] follows from the DPI of $\beta_{1-\bar \gamma}$ by introducing the channel output $Y^n$.
 \item[(b)] follows from the fact that
\begin{equation}
 W_{\mathcal{I}} \rightarrow Y^n \rightarrow \hat W_{\mathcal{I}}
 \end{equation}
 forms a Markov chain under the distribution $p_{W_{\mathcal{I}} ,Y^n, \hat W_{\mathcal{I}} }$.
 \item[(c)] follows from the DPI of $\beta_{1-\bar \gamma}$ by introducing the channel input $X_\mathcal{I}^n$.
 \item[(d)] follows from Definition~\ref{defCode}, which says $X_{T^c}^n$ is a function of $W_{T^c}$.
 \end{enumerate}
 \subsection{Relaxation via Chebyshev's Inequality}
 Following \eqref{eqnBHTFirstChain}, we consider
\begin{equation}
 p_{X_{\mathcal{I}}^n,Y^n, \hat W_\mathcal{I}|W_\mathcal{I}=w_\mathcal{I}}
 \stackrel{\eqref{memorylessStatement}}{=} p_{X_\mathcal{I}^n |W_{\mathcal{I}}=w_\mathcal{I}}p_{\hat W_\mathcal{I}|Y^n}\prod_{k=1}^n p_{Y_k|X_{\mathcal{I},k}}, \label{eqnBHTSecondChain}
\end{equation}
and we obtain from Lemma~\ref{lemmaDPI} and \eqref{eqnBHTSecondChain} that for each $w_\mathcal{I}\in \mathcal{A}$ and each $\xi_{w_\mathcal{I}}>0$,
\begin{align}
&\beta_{1-\bar \gamma}\left(p_{X_\mathcal{I}^n |W_{\mathcal{I}}=w_\mathcal{I}}p_{\hat W_\mathcal{I}|Y^n}\prod_{k=1}^n p_{Y_k|X_{\mathcal{I},k}}\left\|p_{X_\mathcal{I}^n |W_{\mathcal{I}}=w_\mathcal{I}}p_{\hat W_\mathcal{I}|Y^n}\prod_{k=1}^n s_{Y_k|X_{T^c, k}}\right.\right) \notag\\
&\ge\frac{1}{\xi_{w_\mathcal{I}}}\left(1-\bar \gamma - \Pr_{p_{X_{\mathcal{I}}^n,Y^n, \hat W_\mathcal{I}|W_\mathcal{I}=w_\mathcal{I}}}\left\{\prod_{k=1}^n \frac{p_{Y_k|X_{\mathcal{I},k}}(Y_k|X_{\mathcal{I},k})}{s_{Y_{k}|X_{T^c,k}}(Y_k|X_{T^c,k})} \ge \xi_{w_\mathcal{I}} \right\}\right) . \label{eqnBHTThirdChain}
\end{align}
Combining \eqref{eqnBHTReverseChain}, \eqref{eqnBHTFirstChain} and \eqref{eqnBHTThirdChain}, we obtain for each $w_\mathcal{I}\in \mathcal{A}$ and each $\xi_{w_\mathcal{I}}>0$
\begin{equation}
s_{\hat W_T|W_{T^c}}(w_T|w_{T^c}) \ge\frac{1}{\xi_{w_\mathcal{I}}} \left(1-\bar \gamma - \Pr_{p_{X_{\mathcal{I}}^n,Y^n|W_\mathcal{I}=w_\mathcal{I}}}\left\{\prod_{k=1}^n \frac{p_{Y_k|X_{\mathcal{I},k}}(Y_k|X_{\mathcal{I},k})}{s_{Y_{k}|X_{T^c,k}}(Y_k|X_{T^c,k})} \ge \xi_{w_\mathcal{I}} \right\}\right) ,\label{eqnBHTfifthChain}
\end{equation}
which implies that
\begin{align}
&\log\left(\frac{1}{s_{\hat W_T|W_{T^c}}(w_T|w_{T^c})}\right) \notag\\
 & \le \log \xi_{w_\mathcal{I}} - \log\left(1-\bar \gamma - \Pr_{p_{X_{\mathcal{I}}^n,Y^n|W_\mathcal{I}=w_\mathcal{I}}}\left\{\sum_{k=1}^n \log\left( \frac{p_{Y_k|X_{\mathcal{I},k}}(Y_k|X_{\mathcal{I},k})}{s_{Y_{k}|X_{T^c,k}}(Y_k|X_{T^c,k})}\right) \ge \log\xi_{w_\mathcal{I}} \right\}\right).\label{eqnBHTsixthChain}
\end{align}
For each $w_\mathcal{I}\in \mathcal{A}$, let
\begin{align}
\log \xi_{w_\mathcal{I}}&\triangleq \E_{p_{X_{\mathcal{I}}^n,Y^n|W_\mathcal{I}=w_\mathcal{I}}}\left[\sum_{k=1}^n \log\left( \frac{p_{Y_k|X_{\mathcal{I},k}}(Y_k|X_{\mathcal{I},k})}{s_{Y_{k}|X_{T^c,k}}(Y_k|X_{T^c,k})}\right) \right] \notag \\ &
\qquad + \sqrt{\frac{2}{1-\bar \gamma}\Var_{p_{X_{\mathcal{I}}^n,Y^n|W_\mathcal{I}=w_\mathcal{I}}}\left[\sum_{k=1}^n \log\left( \frac{p_{Y_k|X_{\mathcal{I},k}}(Y_k|X_{\mathcal{I},k})}{s_{Y_{k}|X_{T^c,k}}(Y_k|X_{T^c,k})}\right)\right]}\, \, . \label{defXiWI}
\end{align}
Using Chebyshev's inequality, it follows from \eqref{defXiWI} that for each $w_\mathcal{I}\in \mathcal{A}$
\begin{equation}
\Pr_{p_{X_{\mathcal{I}}^n,Y^n|W_\mathcal{I}=w_\mathcal{I}}}\left\{\sum_{k=1}^n \log\left( \frac{p_{Y_k|X_{\mathcal{I},k}}(Y_k|X_{\mathcal{I},k})}{s_{Y_{k}|X_{T^c,k}}(Y_k|X_{T^c,k})}\right) \ge \log\xi_{w_\mathcal{I}} \right\} \le \frac{1-\bar \gamma}{2},
\end{equation}
which implies from \eqref{eqnBHTsixthChain} that
\begin{align}
\log\left(\frac{1}{s_{\hat W_T|W_{T^c}}(w_T|w_{T^c})}\right) \le \log \xi_{w_\mathcal{I}} + \log\left(\frac{2}{1-\bar \gamma}\right).\label{eqnBHTbefore7thChain}
\end{align}
Since $t\mapsto \log\frac{1}{t}$ is convex for $t>0$, by Jensen's inequality
\begin{align}
\sum_{w_{\mathcal{I}}\in \mathcal{A}}p_{W_{\mathcal{I}}} ( w_{\mathcal{I}})\log\left(\frac{1}{s_{\hat W_T|W_{T^c}}(w_T|w_{T^c})}\right)&\ge\log\left(\frac{1}{ \sum_{w_{\mathcal{I}}\in \mathcal{A}}p_{W_\mathcal{I}} ( w_{\mathcal{I}})s_{\hat W_T|W_{T^c}}(w_T|w_{T^c})}\right). \label{eqn:jens*}
\end{align}
We have
\begin{align}
\sum\limits_{w_{\mathcal{I}}\in \mathcal{A}}p_{W_{\mathcal{I}}} ( w_{\mathcal{I}})s_{\hat W_T|W_{T^c}}(w_T|w_{T^c})
& \stackrel{\eqref{eqn:statement(i)}}{=} \frac{1}{|\mathcal{A}|} \sum\limits_{w_{\mathcal{I}}\in \mathcal{A}}s_{\hat W_T|W_{T^c}}(w_T|w_{T^c}) \\
& \stackrel{ \text{(a)} }{=} \frac{1}{|\mathcal{A}|} \sum\limits_{w_T\in \mathcal{A}_T}s_{\hat W_T|W_{T^c}}(w_T|w_{T^c}^*) \\
&\le \frac{1}{|\mathcal{A}|} \sum\limits_{w_T\in \mathcal{W}_T}s_{\hat W_T|W_{T^c}}(w_T|w_{T^c}^*) \\
&= \frac{1}{|\mathcal{A}|} \label{eqn:jens**}
\end{align}
where (a) follows from the definition of $\mathcal{A}_T$ in \eqref{defSetAT} and the fact stated in Statement~(ii) that $w_{T^c}=w_{T^c}^*$ for all $w_\mathcal{I}\in \mathcal{A}$.
Using \eqref{eqn:jens*} and \eqref{eqn:jens**}, we obtain
\begin{align}
\sum_{ w_{\mathcal{I}}\in \mathcal{A}}p_{W_{\mathcal{I}}} ( w_{\mathcal{I}})\log\left(\frac{1}{s_{\hat W_T|W_{T^c}}(w_T|w_{T^c})}\right)&\ge\log|\mathcal{A}|.\label{eqn:jens}
\end{align}
Taking expectation with respect to $p_{W_{\mathcal{I}}}$ on both sides of \eqref{eqnBHTbefore7thChain} and applying~\eqref{eqn:jens}, we obtain
\begin{align}
\log|\mathcal{A}| \le \left(\sum_{w_{\mathcal{I}}\in \mathcal{A}}p_{W_{\mathcal{I}}}(w_{\mathcal{I}})\log \xi_{w_\mathcal{I}}\right) + \log\left(\frac{2}{1-\bar \gamma}\right).\label{eqnBHT7thChain}
\end{align}
 \subsection{Simplification of Log-Likelihood Terms}
In order to simplify \eqref{eqnBHT7thChain}, we will simplify the log-likelihood term in $\log \xi_{w_\mathcal{I}}$ defined in \eqref{defXiWI}. To this end, we first let $x_{i}^n(w_i)\triangleq f_i(w_i)$ ($f_i$ is the encoding function at node~$i$ defined in Definition~\ref{defCode}) and we also let $x_{i,k}(w_i)$ denote the $k^{\text{th}}$ component of $x_{i}^n(w_i)$  for each $i\in\mathcal{I}$ and each $k\in\{1, 2, \ldots n\}$ such that
\begin{equation}
 x_{i}^n(w_i)=(x_{i,1}(w_i),x_{i,2}(w_i), \ldots, x_{i,n}(w_i)).
 \end{equation}
 In addition, we let
\begin{equation}
 x_{\mathcal{I},k}(w_\mathcal{I})\triangleq(x_{1,k}(w_1), x_{2,k}(w_2), \ldots, x_{N,k}(w_N)) ,
 \end{equation}
 and we let
\begin{equation}
x_{T^c,k}(w_{T^c})\triangleq (x_{j,k}(w_j)\,|\, j\in T^c)
\end{equation}
 be a subtuple of $x_{\mathcal{I},k}(w_\mathcal{I})$. Similarly, let
\begin{equation}
 x_\mathcal{I}^n(w_\mathcal{I})\triangleq (x_1^n(w_1), x_2^n(w_2), \ldots, x_N^n(w_N)),
 \end{equation}
 and let
\begin{equation}
 x_{T^c}^n(w_{T^c})\triangleq (x_{j}^n(w_{j}) \,|\, j\in T^c)
 \end{equation}
be a subtuple of $x_\mathcal{I}^n(w_\mathcal{I})$. Using the fact that $X_i^n$ is a function of $W_i$ for all $i\in\mathcal{I}$ and the notations defined above, we obtain from \eqref{defXiWI} that
\begin{align}
\log \xi_{w_\mathcal{I}}&=\E_{p_{Y^n|W_\mathcal{I}=w_\mathcal{I},X_{\mathcal{I}}^n=x_\mathcal{I}^n(w_\mathcal{I})}}\left[\sum_{k=1}^n \log\left( \frac{p_{Y_k|X_{\mathcal{I},k}}(Y_k|x_{\mathcal{I},k}(w_\mathcal{I}))}{s_{Y_{k}|X_{T^c,k}}(Y_k|x_{T^c,k}(w_{T^c}))}\right) \right] \notag \\ &
\qquad + \sqrt{\frac{2}{1-\bar \gamma}\Var_{p_{Y^n|W_\mathcal{I}=w_\mathcal{I},X_{\mathcal{I}}^n=x_\mathcal{I}^n(w_\mathcal{I})}}\left[\sum_{k=1}^n \log\left( \frac{p_{Y_k|X_{\mathcal{I},k}}(Y_k|x_{\mathcal{I},k}(w_\mathcal{I}))}{s_{Y_{k}|X_{T^c,k}}(Y_k|x_{T^c,k}(w_{T^c}))}\right)\right]}\, \, ,
\end{align}
which implies from \eqref{memorylessStatement*} that
\begin{align}
\log \xi_{w_\mathcal{I}}&=\E_{\prod_{k=1}^np_{Y_k|X_{\mathcal{I},k}=x_{\mathcal{I},k}(w_\mathcal{I})} }\left[\sum_{k=1}^n \log\left( \frac{p_{Y_k|X_{\mathcal{I},k}}(Y_k|x_{\mathcal{I},k}(w_\mathcal{I}))}{s_{Y_{k}|X_{T^c,k}}(Y_k|x_{T^c,k}(w_{T^c}))}\right) \right] \notag \\ &
\qquad + \sqrt{\frac{2}{1-\bar \gamma}\Var_{\prod_{k=1}^np_{Y_k|X_{\mathcal{I},k}=x_{\mathcal{I},k}(w_\mathcal{I})} }\left[\sum_{k=1}^n \log\left( \frac{p_{Y_k|X_{\mathcal{I},k}}(Y_k|x_{\mathcal{I},k}(w_\mathcal{I}))}{s_{Y_{k}|X_{T^c,k}}(Y_k|x_{T^c,k}(w_{T^c}))}\right)\right]}\, \, ,
\end{align}
which then implies that
\begin{align}
\log \xi_{w_\mathcal{I}}&=\sum_{k=1}^n\E_{p_{Y_k|X_{\mathcal{I},k}=x_{\mathcal{I},k}(w_\mathcal{I})} }\left[ \log\left( \frac{p_{Y_k|X_{\mathcal{I},k}}(Y_k|x_{\mathcal{I},k}(w_\mathcal{I}))}{s_{Y_{k}|X_{T^c,k}}(Y_k|x_{T^c,k}(w_{T^c}))}\right) \right] \notag \\ &
\qquad + \sqrt{\frac{2}{1-\bar \gamma}\sum_{k=1}^n\Var_{p_{Y_k|X_{\mathcal{I},k}=x_{\mathcal{I},k}(w_\mathcal{I})} }\left[ \log\left( \frac{p_{Y_k|X_{\mathcal{I},k}}(Y_k|x_{\mathcal{I},k}(w_\mathcal{I}))}{s_{Y_{k}|X_{T^c,k}}(Y_k|x_{T^c,k}(w_{T^c}))}\right)\right]}\, \, . \label{defXiWI*}
\end{align}
Following \eqref{defXiWI*}, we use \eqref{defChannelInDefinition*}, \eqref{defChannelInDefinition} and \eqref{defSimulatingDistSKsumRate} to obtain
\begin{align}
&\log\left( \frac{p_{Y_k|X_{\mathcal{I},k}}(Y_k|x_{\mathcal{I},k}(w_\mathcal{I}))}{s_{Y_{k}|X_{T^c,k}}(Y_k|x_{T^c,k}(w_{T^c}))}\right) \notag\\*
& = \frac{1}{2}\log\left(1+\sum_{i\in T}P_i\right)+\frac{\log e}{2(1+\sum_{i\in T}P_i)} \Bigg(-\left(\sum_{i\in T}P_i\right)\left(Y_k-\sum_{i\in\mathcal{I}}x_{i,k}(w_i)\right)^2 \notag \\*
& \qquad + 2\left(\sum_{i\in T}(x_{i,k}(w_i)-\E_{u_{\hat X_{i,k}}}[\hat X_{i,k}])\right)\left(Y_k-\sum_{i\in\mathcal{I}}x_{i,k}(w_i)\right) + \left(\sum_{i\in T}(x_{i,k}(w_i)-\E_{u_{\hat X_{i,k}}}[\hat X_{i,k}])\right)^2\Bigg). \label{eqnBHT8thChain}
\end{align}
For each $w_\mathcal{I}\in \mathcal{A}$ and each $k\in\{1, 2, \ldots, n\}$, it follows from Definition~\ref{defGaussianMAC} that $Y_k-\sum_{i\in\mathcal{I}}x_{i,k}(w_i)$ is a standard normal random variable if $Y_k$ is distributed according to $p_{Y_k|X_{\mathcal{I},k}=x_{\mathcal{I},k}(w_\mathcal{I})}$, which then implies that
\begin{align}
& \E_{p_{Y_k|X_{\mathcal{I},k}=x_{\mathcal{I},k}(w_\mathcal{I})}}\left[\log\left( \frac{p_{Y_k|X_{\mathcal{I},k}}(Y_k|x_{\mathcal{I},k}(w_\mathcal{I}))}{s_{Y_{k}|X_{T^c,k}}(Y_k|x_{T^c,k}(w_{T^c}))}\right)\right] \notag\\*
 &\stackrel{\eqref{eqnBHT8thChain}}{=} \frac{1}{2}\log\left(1+\sum_{i\in T}P_i\right)+\frac{\log e}{2(1+\sum_{i\in T}P_i)}\left(-\left(\sum_{i\in T}P_i\right) + \left(\sum_{i\in T}(x_{i,k}(w_i)-\E_{u_{\hat X_{i,k}}}[\hat X_{i,k}])\right)^2\right) \label{infoSpectrumExpTimek}
\end{align}
and
\begin{align}
&\Var_{p_{Y_k|X_{\mathcal{I},k}=x_{\mathcal{I},k}(w_\mathcal{I})}}\left[\log\left( \frac{p_{Y_k|X_{\mathcal{I},k}}(Y_k|x_{\mathcal{I},k}(w_\mathcal{I}))}{s_{Y_{k}|X_{T^c,k}}(Y_k|x_{T^c,k}(w_{T^c}))}\right)\right] \notag\\*
& \stackrel{\eqref{eqnBHT8thChain}}{=}\Bigg( \frac{\log e}{2(1+\sum_{i\in T}P_i)} \Bigg)^2 \Var_{p_{Y_k|X_{\mathcal{I},k}=x_{\mathcal{I},k}(w_\mathcal{I})}}\Bigg[ -\Bigg(\sum_{i\in T}P_i\Bigg)\Bigg(Y_k-\sum_{i\in\mathcal{I}}x_{i,k}(w_i)\Bigg)^2 \notag\\*
 &\qquad + 2\left(\sum_{i\in T}(x_{i,k}(w_i)-\E_{u_{\hat X_{i,k}}}[\hat X_{i,k}])\right)\Bigg(Y_k-\sum_{i\in\mathcal{I}}x_{i,k}(w_i)\Bigg) \Bigg] \\*
 & = \frac{\left((\sum_{i\in T}P_i)^2 + 2\left(\sum_{i\in T}(x_{i,k}(w_i)-\E_{u_{\hat X_{i,k}}}[\hat X_{i,k}])\right)^2\right)(\log e)^2}{2(1+\sum_{i\in T}P_i)^2}.
 \label{infoSpectrumVarExpTimek}
\end{align}
Define
\begin{equation}
|P_T|\triangleq \sum_{i\in T}P_i \label{defPT}
\end{equation}
and
\begin{equation}
\bar x_{i,k}(w_i)\triangleq x_{i,k}(w_i)-\E_{u_{\hat X_{i,k}}}[\hat X_{i,k}]. \label{defBarXik}
\end{equation}
Combining \eqref{eqnBHT7thChain}, \eqref{defXiWI*}, \eqref{infoSpectrumExpTimek}, \eqref{infoSpectrumVarExpTimek}, \eqref{defPT} and \eqref{defBarXik}, we obtain for each $w_\mathcal{I}\in \mathcal{A}$
\begin{align}
\log|\mathcal{A}| & \le \frac{n}{2}\log\left(1+|P_T|\right)+ \frac{\sum_{w_\mathcal{I}\in \mathcal{A}}p_{W_\mathcal{I}}(w_\mathcal{I})\left(-n|P_T| + \sum_{k=1}^n\left(\sum_{i\in T}\bar x_{i,k}(w_i)\right)^2\right)\log e}{2(1+|P_T|)} \notag\\*
&\quad + \frac{\sum_{w_\mathcal{I}\in \mathcal{A}}p_{W_\mathcal{I}}(w_\mathcal{I})\sqrt{\left(n |P_T|^2 + 2 \sum_{k=1}^n\left(\sum_{i\in T}\bar x_{i,k}(w_i)\right)^2 \right)}\log e}{(1+|P_T|)\sqrt{1-\bar \gamma}} + \log\left(\frac{2}{1-\bar \gamma}\right), 
\end{align}
which implies from Jensen's inequality ($t\mapsto\sqrt{t}$ is concave for $t\ge 0$) that
\begin{align}
\log|\mathcal{A}|
& \le \frac{n}{2}\log\left(1+|P_T|\right)+ \frac{\left(-n|P_T| + \sum_{k=1}^n\sum_{w_\mathcal{I} \in \mathcal{A}}p_{W_\mathcal{I}}(w_\mathcal{I})\left(\sum_{i\in T}\bar x_{i,k}(w_i)\right)^2\right)\log e}{2(1+|P_T|)} \notag\\
&\quad + \frac{\sqrt{n |P_T|^2 + 2 \sum_{k=1}^n\sum_{w_\mathcal{I}\in \mathcal{A}}p_{W_\mathcal{I}}(w_\mathcal{I})\left(\sum_{i\in T}\bar x_{i,k}(w_i)\right)^2 }\log e}{(1+|P_T|)\sqrt{1-\bar \gamma}} + \log\left(\frac{2}{1-\bar \gamma}\right). \label{eqnBHT10thChain}
\end{align}
In the following, we will obtain an upper bound on the crucial term $\sum_{k=1}^n\sum_{w_\mathcal{I}\in \mathcal{A}}p_{W_\mathcal{I}}(w_\mathcal{I})\left(\sum_{i\in T}\bar x_{i,k}(w_i)\right)^2$ which appears in the second and third terms on the right-hand-side of \eqref{eqnBHT10thChain}.
\subsection{Introducing the Quantized Input Distribution to Simplify the Upper Bound}
Following \eqref{eqnBHT10thChain}, we consider for each $k\in\{1, 2, \ldots, n\}$
\begin{align}
&\sum\limits_{w_\mathcal{I}\in \mathcal{A}}p_{W_\mathcal{I}}(w_\mathcal{I})\left(\sum\limits_{i\in T}\bar x_{i,k}(w_i)\right)^2 \notag\\*
& = \sum\limits_{w_T\in \mathcal{A}_T}p_{W_T}(w_T)\left(\sum\limits_{i\in T}\bar x_{i,k}(w_i)\right)^2 \sum\limits_{w_{T^c}\in \mathcal{W}_{T^c}}p_{W_{T^c}|W_T}(w_{T^c}|w_T) \\
& = \sum\limits_{w_T\in \mathcal{A}_T}p_{W_T}(w_T)\left(\sum\limits_{i\in T}\bar x_{i,k}(w_i)\right)^2 \\
& \le \sum\limits_{w_T\in \mathcal{W}_T}p_{W_T}(w_T)\left(\sum\limits_{i\in T}\bar x_{i,k}(w_i)\right)^2. \label{eqnBHT9thChain}
\end{align}
Since $X_i^n$ is a function of $W_i$ for each $i\in T$, it follows from \eqref{defBarXik} that for each $k\in\{1, 2, \ldots, n\}$
\begin{equation}
\sum\limits_{w_T\in \mathcal{W}_T}p_{W_T}(w_T)\left(\sum\limits_{i\in T}\bar x_{i,k}(w_i)\right)^2 = \sum\limits_{x_{T,k}\in \mathcal{X}_T}p_{X_{T,k}}(x_{T,k})\left(\sum\limits_{i\in T} \left(x_{i,k}- \E_{u_{\hat X_{i,k}}}[\hat X_{i,k}]\right)\right)^2,
\end{equation}
which implies from \eqref{eqnBHT9thChain} that
\begin{equation}
\sum\limits_{w_\mathcal{I}\in \mathcal{A}}p_{W_\mathcal{I}}(w_\mathcal{I})\left(\sum\limits_{i\in T}\bar x_{i,k}(w_i)\right)^2 \le \sum\limits_{x_{T,k}\in \mathcal{X}_T}p_{X_{T,k}}(x_{T,k})\left(\sum\limits_{i\in T} \left(x_{i,k}- \E_{u_{\hat X_{i,k}}}[\hat X_{i,k}]\right)\right)^2. \label{eqnBHT10thChain*}
\end{equation}
Recalling the definition of $\hat X_T^n$ and $\hat{\mathcal{X}}_T^n$ in \eqref{defQuantizedDistributionInMainProof} and \eqref{eqn:quant_alpha} respectively,
 we write for each $k\in\{1, 2, \ldots, n\}$
\begin{align}
&\sum\limits_{x_{T,k}\in \mathcal{X}_T}p_{X_{T,k}}(x_{T,k})\left(\sum\limits_{i\in T} \left(x_{i,k}- \E_{u_{\hat X_{i,k}}}[\hat X_{i,k}]\right)\right)^2 \notag\\*
& = \sum\limits_{x_{T,k}\in \mathcal{X}_T, \hat x_{T,k}\in \hat{\mathcal{X}}_T}p_{X_{T,k}, \hat X_{T,k}}(x_{T,k}, \hat x_{T,k})\left(\sum\limits_{i\in T} \left(x_{i,k}-\hat x_{i,k}+\hat x_{i,k}- \E_{u_{\hat X_{i,k}}}[\hat X_{i,k}]\right)\right)^2 \\
&= \sum\limits_{x_{T,k}\in \mathcal{X}_T, \hat x_{T,k}\in \hat{\mathcal{X}}_T}p_{X_{T,k}, \hat X_{T,k}}(x_{T,k}, \hat x_{T,k})\left(\sum\limits_{i\in T} (x_{i,k}-\hat x_{i,k})\right)^2 \notag\\*
&\qquad +2\sum\limits_{x_{T,k}\in \mathcal{X}_T, \hat x_{T,k}\in \hat{\mathcal{X}}_T}p_{X_{T,k}, \hat X_{T,k}}(x_{T,k}, \hat x_{T,k})\left(\sum\limits_{i\in T} (x_{i,k}-\hat x_{i,k})\right)\left(\sum\limits_{i\in T}(\hat x_{i,k}- \E_{u_{\hat X_{i,k}}}[\hat X_{i,k}])\right)\notag\\*
&\qquad +\sum\limits_{\hat x_{T,k}\in \hat{\mathcal{X}}_T}p_{\hat X_{T,k}}(\hat x_{T,k})\left(\sum\limits_{i\in T}(\hat x_{i,k}- \E_{u_{\hat X_{i,k}}}[\hat X_{i,k}])\right)^2 \\
&\le \sum\limits_{x_{T,k}\in \mathcal{X}_T, \hat x_{T,k}\in \hat{\mathcal{X}}_T}p_{X_{T,k}, \hat X_{T,k}}(x_{T,k}, \hat x_{T,k})\left|\sum\limits_{i\in T} (x_{i,k}-\hat x_{i,k})\right|^2 \notag\\*
&\qquad +2\sum\limits_{x_{T,k}\in \mathcal{X}_T, \hat x_{T,k}\in \hat{\mathcal{X}}_T}p_{X_{T,k}, \hat X_{T,k}}(x_{T,k}, \hat x_{T,k})\left|\sum\limits_{i\in T} (x_{i,k}-\hat x_{i,k})\right|\left|\sum\limits_{i\in T}(\hat x_{i,k}- \E_{u_{\hat X_{i,k}}}[\hat X_{i,k}])\right|\notag\\*
&\qquad +\sum\limits_{\hat x_{T,k}\in \hat{\mathcal{X}}_T}p_{\hat X_{T,k}}(\hat x_{T,k})\left(\sum\limits_{i\in T}(\hat x_{i,k}- \E_{u_{\hat X_{i,k}}}[\hat X_{i,k}])\right)^2 \\
&\le \sum\limits_{x_{T,k}\in \mathcal{X}_T, \hat x_{T,k}\in \hat{\mathcal{X}}_T}p_{X_{T,k}, \hat X_{T,k}}(x_{T,k}, \hat x_{T,k})\left(\sum\limits_{i\in T} |x_{i,k}-\hat x_{i,k}|\right)^2 \notag\\*
&\qquad +2\sum\limits_{x_{T,k}\in \mathcal{X}_T, \hat x_{T,k}\in \hat{\mathcal{X}}_T}p_{X_{T,k}, \hat X_{T,k}}(x_{T,k}, \hat x_{T,k})\left(\sum\limits_{i\in T} |x_{i,k}-\hat x_{i,k}|\right)\left(\sum\limits_{i\in T}(|\hat x_{i,k}| + \E_{u_{\hat X_{i,k}}}[|\hat X_{i,k}|])\right)\notag\\*
&\qquad +\sum\limits_{\hat x_{T,k}\in \hat{\mathcal{X}}_T}p_{\hat X_{T,k}}(\hat x_{T,k})\left(\sum\limits_{i\in T}(\hat x_{i,k}- \E_{u_{\hat X_{i,k}}}[\hat X_{i,k}])\right)^2 \\
&\stackrel{\text{(a)}}{\le}\frac{|T|^2}{n^2} +\frac{4|T|}{\sqrt{n}}\left(\sum_{i\in T}\sqrt{P_i}\right) +\sum\limits_{\hat x_{T,k}\in \hat{\mathcal{X}}_T}p_{\hat X_{T,k}}(\hat x_{T,k})\left(\sum\limits_{i\in T}(\hat x_{i,k}- \E_{u_{\hat X_{i,k}}}[\hat X_{i,k}])\right)^2 \label{eqnBHT11thChain}
\end{align}
where (a) follows from the facts below for each $i\in T$, each $k\in\{1, 2, \ldots, n\}$ and each $x_{i,k}\in \mathcal{X}_i$ (recall the definition of $\hat x_{i,k}$ in \eqref{defQuantizedDistributionInMainProof}):
\begin{equation}
|x_{i,k}-\hat x_{i,k}|\stackrel{\eqref{defOmega*}}{\le} \frac{1}{n} \label{boundOnDifference}
\end{equation}
and
\begin{equation}
|\hat x_{i,k}| \stackrel{\eqref{defOmega}}{\le} |x_{i,k}| \stackrel{\eqref{powerConstraint}}{\le} \sqrt{n P_i} \, . \label{boundOnHatXik}
\end{equation}
\subsection{Approximating the Quantized Input Distribution by a Product Distribution}
In order to bound the last term in \eqref{eqnBHT11thChain}, we use the bound in \eqref{eqn:statement(iv)} for bounding $p_{\hat X_{T,k}}(\hat x_{T,k})$ in terms of $u_{\hat X_{T,k}}(\hat x_{T,k})$ to obtain
\begin{align}
&\sum\limits_{\hat x_{T,k}\in \hat{\mathcal{X}}_T}p_{\hat X_{T,k}}(\hat x_{T,k})\left(\sum\limits_{i\in T}(\hat x_{i,k}- \E_{u_{\hat X_{i,k}}}[\hat X_{i,k}])\right)^2 \notag\\
& \quad \le \sum\limits_{\hat x_{T,k} \in \hat{\mathcal{X}}_T}\left(\left(1+\sqrt{\frac{\log n}{n}}\right)\prod_{i\in T}u_{\hat X_{i,k}}(\hat x_{i,k})+ \frac{1}{n^{4|T|}}\right)\left(\sum\limits_{i\in T}(\hat x_{i,k}- \E_{u_{\hat X_{i,k}}}[\hat X_{i,k}])\right)^2 \label{eqnBHT12thChain_a}\\
& \quad = \sum\limits_{\hat x_{T,k} \in \hat{\mathcal{X}}_T}\Bigg[\left(1+\sqrt{\frac{\log n}{n}}\right)\prod_{i\in T}u_{\hat X_{i,k}}(\hat x_{i,k})\left(\sum\limits_{i\in T}(\hat x_{i,k}- \E_{u_{\hat X_{i,k}}}[\hat X_{i,k}])\right)^2 \notag\\*
&\qquad\qquad\qquad\qquad+ \frac{1}{n^{4|T|}} \left(\sum\limits_{i\in T}(\hat x_{i,k}- \E_{u_{\hat X_{i,k}}}[\hat X_{i,k}])\right)^2 \Bigg]\label{eqnBHT12thChain}
\end{align}
 for each $k\in\{1, 2, \ldots, n\}$.
The bound in~\eqref{eqnBHT12thChain} consists of two distinct terms which we now bound separately. Consider the following two chains of inequalities for each $k\in\{1, 2, \ldots, n\}$:
\begin{align}
&\sum\limits_{\hat x_{T,k} \in \hat{\mathcal{X}}_T} \left(\prod_{i\in T}u_{\hat X_{i,k}}(\hat x_{i,k})\right)\left(\sum\limits_{i\in T}(\hat x_{i,k}- \E_{u_{\hat X_{i,k}}}[\hat X_{i,k}])\right)^2 \notag \\*
& = \sum_{i\in T} \E_{u_{\hat X_{i,k}}}\left[ (\hat X_{i,k}- \E_{u_{\hat X_{i,k}}}[\hat X_{i,k}])^2\right] \\*
&\le \sum_{i\in T} \E_{u_{\hat X_{i,k}}} \left[\hat X_{i,k}^2\right] \label{eqnBHT13thChain}
\end{align}
and
\begin{align}
&\sum\limits_{\hat x_{T,k} \in \hat{\mathcal{X}}_T}\left(\sum\limits_{i\in T}(\hat x_{i,k}- \E_{u_{\hat X_{i,k}}}[\hat X_{i,k}])\right)^2 \notag\\*
&\quad \le \sum\limits_{\hat x_{T,k} \in \hat{\mathcal{X}}_T} \left(|T|\max\limits_{i\in T}\left\{|\hat x_{i,k}- \E_{u_{\hat X_{i,k}}}[\hat X_{i,k}]|\right\} \right)^2 \\
&\quad = |T|^2\sum\limits_{\hat x_{T,k} \in \hat{\mathcal{X}}_T} \max\limits_{i\in T}\left\{(\hat x_{i,k}- \E_{u_{\hat X_{i,k}}}[\hat X_{i,k}])^2\right\} \\
&\quad \le |T|^2\sum\limits_{\hat x_{T,k} \in \hat{\mathcal{X}}_T} \sum_{i\in T}(\hat x_{i,k}- \E_{u_{\hat X_{i,k}}}[\hat X_{i,k}])^2 \\
&\quad \stackrel{\text{(a)}}{\le} 2|T|^2\sum\limits_{\hat x_{T,k} \in \hat{\mathcal{X}}_T} \sum_{i\in T}\left(\hat x_{i,k}^2 + (\E_{u_{\hat X_{i,k}}}[\hat X_{i,k}])^2\right) \\
&\quad \stackrel{\eqref{boundOnHatXik}}{\le} 2|T|^2\sum\limits_{\hat x_{T,k} \in \hat{\mathcal{X}}_T} \sum_{i\in T}2nP_i \\
&\quad \stackrel{\text{(b)}}{\le} 4n|T|^2 |P_T| | \hat{\mathcal{X}}_T|\\
&\quad \stackrel{\eqref{eqn:size_Xhat}}{<} 4n^{3|T|}|T|^2 |P_T|\prod\limits_{i \in T}(2\sqrt{P_i} + 3), \label{eqnBHT14thChain}
\end{align}
where
\begin{enumerate}
\item[(a)] follows from the fact that $(a-b)^2\le 2a^2 + 2b^2$ for all real numbers $a$ and $b$.
\item[(b)] follows from the definition of $|P_T|$ in~\eqref{defPT}.
\end{enumerate}
Combining \eqref{eqnBHT12thChain}, \eqref{eqnBHT13thChain} and \eqref{eqnBHT14thChain}, we obtain for each $k\in\{1, 2, \ldots, n\}$
\begin{align}
&\sum_{\hat x_{T,k} \in \hat{\mathcal{X}}_T}p_{\hat X_{T,k}}(\hat x_{T,k})\left(\sum_{i\in T}(\hat x_{i,k}- \E_{u_{\hat X_{i,k}}}[\hat X_{i,k}])\right)^2 \notag\\*
&\quad \le \left(1+\sqrt{\frac{\log n}{n}}\right)\sum_{i\in T} \E_{u_{\hat X_{i}}} \left[\hat X_{i,k}^2\right] + 4n^{-|T|}|T|^2 |P_T|\prod\limits_{i \in T}(2\sqrt{P_i} + 3),
\end{align}
which implies from \eqref{eqnBHT10thChain*} and \eqref{eqnBHT11thChain} that
\begin{align}
&\sum\limits_{w_\mathcal{I}\in \mathcal{A}}p_{W_\mathcal{I}}(w_\mathcal{I})\left(\sum\limits_{i\in T}\bar x_{i,k}(w_i)\right)^2 \notag\\
&\quad \le \frac{|T|^2}{n^2} +\frac{4|T|}{\sqrt{n}}\left(\sum_{i\in T}\sqrt{P_i}\right) + \left(1+\sqrt{\frac{\log n}{n}}\right)\sum_{i\in T} \E_{u_{\hat X_{i}}} \left[\hat X_{i,k}^2\right] + 4n^{-|T|}|T|^2 |P_T|\prod\limits_{i \in T}(2\sqrt{P_i} + 3). \label{eqnBHT15thChain}
\end{align}
Using \eqref{eqnBHT15thChain} and \eqref{CorollaryWringingSt5} and recalling that $|T|\ge 1$ (because $T$ is non-empty), we obtain
\begin{align}
&\sum_{k=1}^n\sum\limits_{w_\mathcal{I}\in \mathcal{A}}p_{W_\mathcal{I}}(w_\mathcal{I})\left(\sum\limits_{i\in T}\bar x_{i,k}(w_i)\right)^2 \notag\\*
&\quad \le n|P_T| + \sqrt{n\log n}|P_T| + 4\sqrt{n}|T|\left(\sum_{i\in T}\sqrt{P_i}\right) + 4|T|^2 |P_T|\prod\limits_{i \in T}(2\sqrt{P_i} + 3) + \frac{|T|^2}{n}. \label{eqnBHT16thChain}
\end{align}
To simplify notation, let
\begin{equation}
\kappa_1\triangleq 4|T|\left(\sum_{i\in T}\sqrt{P_i}\right)\,\quad\mbox{and}\quad\kappa_2\triangleq 4|T|^2 |P_T|\prod\limits_{i \in T}(2\sqrt{P_i} + 3) \label{eqnBHT16thChain*}
\end{equation}
 be two constants that are independent of~$n$. Then, we combine \eqref{eqnBHT10thChain} and \eqref{eqnBHT16thChain} to yield
\begin{align}
\log|\mathcal{A}|
& \le \frac{n}{2}\log\left(1+|P_T|\right)+ \frac{\left(\sqrt{n\log n}|P_T| + \sqrt{n}\kappa_1 + \kappa_2 + n^{-1}|T|^2\right)\log e}{2(1+|P_T|)} \notag\\*
&\quad + \frac{\sqrt{n |P_T|( |P_T|+ 2) + 2\sqrt{n \log n}|P_T| + 2\sqrt{n}\kappa_1 + 2\kappa_2 + 2n^{-1}|T|^2}\log e}{(1+|P_T|)\sqrt{1-\bar \gamma}} + \log\left(\frac{2}{1-\bar \gamma}\right). \label{eqnBHT16thChaina}
\end{align}
Combining \eqref{eqn:statement(ii)} and \eqref{eqnBHT16thChaina}, we obtain
\begin{align}
& \left(\frac{-4|T|(1+3\bar \gamma)}{1-\bar \gamma}\right)\sqrt{n\log n} + \log\left(\frac{1-\bar \gamma}{2(1+\bar \gamma)}\right) + \sum_{i\in T}\log M_i^{(n)} \notag\\
&\quad \le \frac{n}{2}\log\left(1+|P_T|\right)+ \frac{\left(\sqrt{n\log n}|P_T| + \sqrt{n}\kappa_1 + \kappa_2 + n^{-1}|T|^2\right)\log e}{2(1+|P_T|)} \notag\\*
&\qquad \quad + \frac{\sqrt{n |P_T|( |P_T|+ 2) + 2\sqrt{n \log n}|P_T| + 2\sqrt{n}\kappa_1 + 2\kappa_2 + 2n^{-1}|T|^2}\log e}{(1+|P_T|)\sqrt{1-\bar \gamma}} + \log\left(\frac{2}{1-\bar \gamma}\right). \label{eqnBHT17thChain}
\end{align}
Dividing both sides of \eqref{eqnBHT17thChain} by $n$ and taking limit inferior as~$n$ goes to infinity, we obtain from \eqref{achievableRateInProof} that
\eqref{goalInProof} holds as desired. This completes the proof of Theorem~\ref{thmMainResult}.

\subsection{Discussion on the Choices of the Quantizer's Precision and the Parameters Used in the Wringing Technique in~\eqref{identification1}} \label{sectionDiscussionQuantizer}
Our choice of~$\delta$ in~\eqref{identification1} has been optimized in the following sense. If $\delta$ is chosen such that $\delta = o\left(\sqrt{\frac{\log n}{n}}\right)$, then the second-order term on the RHS of~\eqref{eqnBHT17thChain} would be $\omega\left(\sqrt{n\log n}\right)$ (cf.\ \eqref{eqn:statement(iv)} and~\eqref{eqnBHT12thChain_a}), which then leads to an upper bound on~$\sum_{i\in T}\log M_i^{(n)}$ with a looser (larger) second-order term $\omega(\sqrt{n\log n})$; if $\delta$ is chosen such that $\delta = \omega\left(\sqrt{\frac{\log n}{n}}\right)$, then the magnitude of the first term on the LHS of~\eqref{eqnBHT17thChain} would be $\omega\left(\sqrt{n\log n}\right)$ (cf.\ \eqref{eqn:statement(ii)}), which then leads to an upper bound on~$\sum_{i\in T}\log M_i^{(n)}$ with a looser second-order term $\omega(\sqrt{n\log n})$. Hence our choice of $\delta=\sqrt{\frac{\log n}{n}}$ ``balances" the rates of growth of the two second-order terms in~\eqref{eqnBHT17thChain}. In this sense, our choice of $\delta$ is optimal.

We now discuss the choice of the quantizer's precision $\Delta_n=1/n$ as shown in \eqref{eqn:quant_alpha}. Based on this choice of $\Delta_n$, we note that any choice of $\lambda$ in~\eqref{identification1} satisfying $\lambda n^{3|T|+1}=o(\sqrt{n\log n})$ does not affect the second-order term of the resultant upper bound on $\sum_{i\in T}\log M_i^{(n)}$ implied by~\eqref{eqnBHT17thChain}. In particular, the current choice $\lambda=\frac{1}{n^{4|T|}}$ stated in~\eqref{identification1} leads to the rightmost term in~\eqref{eqn:statement(iv)}, which contributes to the fourth constant term in~\eqref{eqnBHT16thChain} as well as the constant term on the RHS of~\eqref{eqnBHT17thChain}.

If the quantizer's precision is chosen to be some other $\Delta_n^\prime$, then it can be seen by inspecting~\eqref{boundOnDifference}, the upper bound obtained at step~(a) in the chain of inequalities leading to \eqref{eqnBHT11thChain}, \eqref{eqnBHT16thChain} and~\eqref{eqnBHT16thChaina} that the second-order term of resultant upper bound on $\sum_{i\in T}\log M_i^{(n)}$ is $\Omega\left(\max\{\sqrt{n\log n}, \Delta_n^\prime n^{3/2}\}\right)$. In particular, if $\Delta_n^\prime$ is chosen such that $\Omega(\frac{1}{n^a}) \le   \Delta_n^\prime\le O\left(\frac{\sqrt{\log n}}{n}\right) $ for any fixed $a\ge 1$, we can follow similar calculations (with a slight modification of $\lambda$) to conclude that the second-order term of the upper bound on $\sum_{i\in T}\log M_i^{(n)}$ is proportional to $\sqrt{n\log n}$. As explained in the second remark after Lemma~\ref{lemmaWringing}, as long as $\Delta_n^\prime$ decays to zero no faster than polynomially in~$n$, then $|\hat{\mathcal{X}}_T|$ grows at most polynomially fast in~$n$, which will ensure that the asymptotic rates of the resultant sequence of subcodes obtained from the wringing step are the same as that of the original sequence of codes. However, if $\Delta_n^\prime$ decays to zero exponentially fast (i.e., $\Delta_n^\prime = O(2^{-nb})$ for some $b>0$), then $|\hat{\mathcal{X}}_T|$ will grow exponentially fast in~$n$  and the RHS of~\eqref{eqn:statement(II)} will decay exponentially rather than polynomially fast. This in turn causes the asymptotic rates of the resultant sequence of subcodes to decrease by a positive quantity, thus resulting in a loose first-order term on the RHS of the final inequality~\eqref{eqnBHT17thChain} (which does not match the corresponding term in the Cover-Wyner capacity region). Therefore, with this choice of $\Delta_n^\prime$, the strong converse cannot be shown.

%
%


\section{Interference Channel under Strong Interference Regime}\label{sectionIC}
 The capacity region of a two-source two-destination Gaussian interference channel (IC) under strong interference was derived by Han and Kobayashi~\cite{Han81} and Sato~\cite{Sato78}. Let $P_1,P_2$ be the received signal-to-noise ratios and let $I_1, I_2$ be the received interference-to-noise ratios~\cite[Sec.~6.4]{elgamal}. Under the formulation of the Gaussian IC under {\em strong interference}, it is assumed that $I_2\ge P_1$ and $I_1\ge P_2$. Under this condition, the capacity region was shown in \cite[Th.~5.2]{Han81} to be the Han-Kobayashi region
\begin{equation}
\mathcal{R}_{\text{\tiny HK-S}} \triangleq \left\{(R_1, R_2)\in \mathbb{R}_+^2 \left| \: \parbox[c]{3.6 in}{$R_1\le \frac{1}{2}\log(1+P_1)$, \vspace{0.04 in}\\ $R_2\le \frac{1}{2}\log(1+P_2) $, \vspace{0.04 in} \\ $R_1+R_2\le \min\{\frac{1}{2}\log(1+P_1+I_1),\frac{1}{2}\log(1+P_2+I_2) \} $} \right.\right\}. \label{strong_interference}
\end{equation}
 By applying Theorem~\ref{thmMainResult} to each of the decoders of the two-source two-destination Gaussian IC, we can show that the corresponding $(\varepsilon_1, \varepsilon_2)$-capacity region $\mathcal{C}_{\varepsilon_1, \varepsilon_2}$ is outer bounded as
\begin{equation}
 \mathcal{C}_{\varepsilon_1, \varepsilon_2} \subseteq \mathcal{R}_{\text{\tiny HK-S}}
 \end{equation}
 as long as $\varepsilon_1+\varepsilon_2<1$, where $\varepsilon_i$ characterizes the asymptotic average probability of destination~$i$ decoding message~$i$ wrongly.
Since the rate pairs in $\mathcal{R}_{\text{\tiny HK-S}} $ are $(0, 0)$-achievable via simultaneous non-unique decoding~\cite[Sec.~6.4]{elgamal},  we have
\begin{equation}
 \mathcal{C}_{\varepsilon_1, \varepsilon_2} =\mathcal{R}_{\text{\tiny HK-S}}
 \end{equation}
 as long as $\varepsilon_1+\varepsilon_2<1$.
 The strong converse (in fact, the complete second-order asymptotics) for the Gaussian IC under the more restrictive condition of strictly very strong interference was shown by Le, Tan, and Motani~\cite{LTM14}. In the rest of this section, we will describe the formulation of the Gaussian IC under strong interference and present in Section~\ref{sectionProofOfThmIC} the corresponding strong converse result.
 \subsection{Problem Formulation and Main Result}
 We follow the standard setting of the Gaussian IC under strong interference as given in \cite[Sec.~V]{Han81}.
 The Gaussian IC under strong interference consists of two sources, denoted by~$\mathrm{s}_1$ and $\mathrm{s}_2$ respectively, and two destinations, denoted by~$\mathrm{d}_1$ and $\mathrm{d}_2$ respectively. For each $i\in\{1,2\}$, $\mathrm{s}_i$ chooses a message~$W_i$ and transmits~$X_i^n$ in~$n$ time slots, and $\mathrm{d}_i$ receives $Y_i^n$ in~$n$ time slots and declares~$\hat W_i$ to be the transmitted $W_i$. The channel law in each time slot~$k$ is
 \begin{equation}
\left[\begin{array}{c}Y_{1,k} \\ Y_{2,k}\end{array}\right] = \left[\begin{array}{cc}1 & g_{12}\\ g_{21} & 1\end{array}\right]\left[\begin{array}{c}X_{1,k} \\ X_{2,k}\end{array}\right] + \left[\begin{array}{c}Z_{1,k} \\ Z_{2,k}\end{array}\right], \label{channelLawIC}
 \end{equation}
 where $g_{21}$ and $g_{12}$ are two real constants characterizing the channel gains of the interference links, and $\left\{(Z_{1,k}, Z_{2,k})\right\}_{k=1}^n$ are $n$ independent copies of a Gaussian random vector denoted by $(Z_1, Z_2)$ ($Z_1$ and $Z_2$ need not be independent) such that
 \begin{align}
 \E\left[Z_{1}\right]=\E\left[Z_{2}\right] = 0 \label{channelLawIC1}
 \end{align}
 and
  \begin{align}
 \E\left[Z_{1}^2\right]=\E\left[Z_{2}^2\right] = 1.  \label{channelLawIC2}
 \end{align}
For each $i\in\{1,2\}$, the codewords transmitted by~$\mathrm{s}_i$ should satisfy the peak power constraint
 \begin{equation}
 \Pr\left\{\|X_i^n\|^2 \le nP_i\right\}=1
 \end{equation}
 for some $P_i>0$.
 We assume that the IC is under strong interference, i.e., $g_{12}^2\ge 1$ and $g_{21}^2\ge 1$, which implies that
 \begin{align}
I_1\triangleq g_{12}^2P_2 \ge P_2 \label{defI1}
 \end{align}
 and
  \begin{align}
I_2\triangleq  g_{21}^2P_1 \ge P_1, \label{defI2}
 \end{align}
 where $I_1$ and $I_2$ characterize the interference power received at~$\mathrm{d}_1$ and $\mathrm{d}_2$ respectively (cf.\ \eqref{channelLawIC}). The Gaussian IC is characterized by some conditional probability density function $q_{Y_1, Y_2|X_1, X_2}$ and we define the Gaussian IC in a similar way to a Gaussian MAC (cf.\ Definition~\ref{defGaussianMAC}) such that \eqref{channelLawIC}, \eqref{channelLawIC1} and \eqref{channelLawIC2} hold. In addition, we define a length-$n$ code for the Gaussian IC as follows.
 \medskip
 \begin{Definition}\label{defCodeIC}
An {\em $(n, M_1^{(n)}, M_2^{(n)}, P_1, P_2)$-code} for the Gaussian IC consists of the
following:
\begin{enumerate}
\item A message set
$
\mathcal{W}_{i}\triangleq \{1, 2, \ldots, M_i^{(n)}\}
$
 at node~$i$ for each $i\in \{1,2\}$, where
$W_i$ is uniform on $\mathcal{W}_{i}$.

\item An encoding function
$
f_i : \mathcal{W}_{i}\rightarrow \mathbb{R}^n
$
 for each $i\in \{1,2\}$, where $f_i$ is the encoding function at node~$i$ such that
$
X_{i}^n=f_i(W_{i})
$
and
$
\|f_i(w_{i})\|^2 \le nP_i \label{powerConstraintIC}
$
for all $w_i\in\mathcal{W}_i$. 
\item A (possibly stochastic) decoding function
$
\varphi_i:
\mathbb{R}^{n} \rightarrow \mathcal{W}_i
$
for each $i\in\{1,2\}$,
where $\varphi_i$ is used by node~$\mathrm{d}_i$ to estimate~$W_i$, i.e.,
$ \hat W_i = \varphi_i(Y_i^{n}).$
\end{enumerate}
 \end{Definition}
 \medskip

We define an $(n, M_1^{(n)}, M_2^{(n)}, P_1, P_2, \varepsilon_1, \varepsilon_2)_{\text{avg}}$-code as follows. \medskip
 \begin{Definition}\label{defErrorIC}
 For an $(n, M_1^{(n)}, M_2^{(n)}, P_1, P_2)$-code defined on the Gaussian IC, the {\em average probability of decoding error for $W_i$} is defined for each $i\in\{1,2\}$ as
\begin{equation}
\Pr\big\{\hat W_i \ne W_i\big\}.
\end{equation}
An $(n, M_1^{(n)}, M_2^{(n)}, P_1, P_2)$-code with $\Pr\big\{\hat W_1 \ne W_1\big\}\le \varepsilon_1$ and $\Pr\big\{\hat W_2 \ne W_2\big\}\le \varepsilon_2$ is called an $(n, M_1^{(n)}, M_2^{(n)}, \linebreak P_1, P_2, \varepsilon_1, \varepsilon_2)_{\text{avg}}$-code.
 \end{Definition}
 \medskip

 For each $\varepsilon_1\in[0,1)$ and each $\varepsilon_2\in[0,1)$, we define an $(\varepsilon_1, \varepsilon_2)$-achievable rate pair as in Definition~\ref{defAchievableRate}, and we define
 the $(\varepsilon_1, \varepsilon_2)$-capacity region, denoted by $\mathcal{C}_{\varepsilon_1, \varepsilon_2}$, to be the set of $(\varepsilon_1, \varepsilon_2)$-achievable rate pairs.
 The following theorem is the main result in this section.
 \medskip
 \begin{Theorem}\label{thmMainResultIC}
For each $\varepsilon_1\in[0,1)$ and each $\varepsilon_2\in[0,1)$ such that $\varepsilon_1+\varepsilon_2<1$,
\begin{equation}
\mathcal{C}_{\varepsilon_1, \varepsilon_2} = \mathcal{R}_{\text{\tiny HK-S}}.
\end{equation}
\end{Theorem}
 \subsection{Proof of Theorem~\ref{thmMainResultIC}} \label{sectionProofOfThmIC}
We need the following definitions and lemma before presenting the proof of Theorem~\ref{thmMainResultIC}. The definition below concerning a multicast code differs from Definition~\ref{defCodeIC} in the decoding functions only, but we state the whole definition for clarity. Essentially, a multicast code for the Gaussian IC is the same as a standard code except that each decoder must output estimates of \emph{both} messages.
\medskip
 \begin{Definition}\label{defCodeICandMAC}
An {\em $(n, M_1^{(n)}, M_2^{(n)}, P_1, P_2)$-multicast code} for the Gaussian IC consists of the
following:
\begin{enumerate}
\item A message set
$
\mathcal{W}_{i}\triangleq \{1, 2, \ldots, M_i^{(n)}\}
$
 at node~$i$ for each $i\in \{1,2\}$, where
$W_i$ is uniform on $\mathcal{W}_{i}$.

\item An encoding function
$
f_i : \mathcal{W}_{i}\rightarrow \mathbb{R}^n
$
 for each $i\in \{1,2\}$, where $f_i$ is the encoding function at node~$i$ such that
$
X_{i}^n=f_i(W_{i})
$
and
$
\|f_i(w_{i})\|^2 \le nP_i 
$
for all $w_i\in\mathcal{W}_i$. 
\item A (possibly stochastic) decoding function
$
\varphi_i:
\mathbb{R}^{n} \rightarrow \mathcal{W}_1\times \mathcal{W}_2
$
for each $i\in\{1,2\}$,
where $\varphi_i$ is used by node~$\mathrm{d}_i$ to estimate both $W_1$ and $W_2$ such that the pair of message estimates is
$
(\hat W_{1, \mathrm{d}_i}, \hat W_{2, \mathrm{d}_i}) \triangleq \varphi_i(Y_i^{n})
$.
\end{enumerate}
 \end{Definition}
 \medskip

We define an $(n, M_1^{(n)}, M_2^{(n)}, P_1, P_2, \varepsilon_1, \varepsilon_2)_{\text{avg}}$-multicast code as follows. Note that the multicast code is used for the Gaussian IC but not a general multicast channel. \medskip
 \begin{Definition}\label{defErrorICandMAC}
 For an $(n, M_1^{(n)}, M_2^{(n)}, P_1, P_2)$-multicast code defined on the Gaussian IC, the {\em average probability of decoding error at destination $\mathrm{d}_i$} is defined for each $i\in\{1,2\}$ as
\begin{equation}
\Pr\left\{\left\{\hat W_{1, \mathrm{d}_i} \ne W_1
\right\} \cup \left\{ \hat W_{2, \mathrm{d}_i} \ne W_2\right\}\right\}.
\end{equation}
An $(n, M_1^{(n)}, M_2^{(n)}, P_1, P_2)$-multicast code with average probability of decoding error at destination $\mathrm{d}_i$ no larger than $\varepsilon_i$ for each $i\in\{1,2\}$ is called an $(n, M_1^{(n)}, M_2^{(n)}, P_1, P_2, \varepsilon_1, \varepsilon_2)_{\text{avg}}$-code.
 \end{Definition}
\medskip

The following lemma plays a crucial role in extending our strong converse result for the Gaussian MAC to the Gaussian IC under strong interference, because it relates the error probabilities for standard codes defined for the Gaussian IC in Definition~\ref{defErrorIC} to the error probabilities for multicast-codes defined for the Gaussian IC in Definition~\ref{defErrorICandMAC}.
\medskip
\begin{Lemma}\label{lemmaIC}
For each $(n, M_1^{(n)}, M_2^{(n)}, P_1, P_2, \varepsilon_1, \varepsilon_2)_{\text{avg}}$-code for the Gaussian IC, there exists an $(n, M_1^{(n)}, M_2^{(n)}, \linebreak P_1, P_2, \varepsilon_1+\varepsilon_2, \varepsilon_1+\varepsilon_2)_{\text{avg}}$-multicast code for the Gaussian IC.
\end{Lemma}
\begin{IEEEproof}
Suppose we are given an $(n, M_1^{(n)}, M_2^{(n)}, P_1, P_2, \varepsilon_1, \varepsilon_2)_{\text{avg}}$-code whose encoding and stochastic decoding functions are denoted by $(f_1, f_2)$ and $(\varphi_1, \varphi_2)$ respectively (cf.\ Definition~\ref{defCodeIC}). Let $p_{W_1, W_2, X_1^n, X_2^n, Y_1^n, Y_2^n, Z_1^n, Z_2^n}$ be the probability distribution induced by the $(n, M_1^{(n)}, M_2^{(n)}, P_1, P_2, \varepsilon_1, \varepsilon_2)_{\text{avg}}$-code. By Definition~\ref{defErrorIC}, we have for each $i\in\{1,2\}$
\begin{equation}
\Pr_{p_{W_i, Y_i^n}}\left\{\varphi_i(Y_i^n)\ne W_i\right\} \le \varepsilon_i\, , \label{eqn0LemmaIC}
\end{equation}
which implies from \eqref{channelLawIC} that
\begin{equation}
\Pr_{p_{W_1, W_2, Z_1^n}}\left\{\varphi_1(f_1(W_1) + g_{12}f_2(W_2)+Z_1^n)\ne W_1\right\}\le \varepsilon_1 \label{eqn1LemmaIC}
\end{equation}
and
\begin{equation}
\Pr_{p_{W_1, W_2, Z_2^n}}\left\{\varphi_2(g_{21} f_1(W_1) + f_2(W_2)+Z_2^n)\ne W_2\right\}\le \varepsilon_2\, . \label{eqn2LemmaIC}
\end{equation}
In the rest of the proof, we construct new stochastic decoding functions at~$\mathrm{d}_1$ and~$\mathrm{d}_2$, denoted by $\varphi_1^\prime$ and $\varphi_2^\prime$ respectively, such that $(\varphi_1, \varphi_1^\prime)$ and $(\varphi_2, \varphi_2^\prime)$ can be viewed as the stochastic decoding functions of an $(n, M_1^{(n)}, M_2^{(n)},\linebreak P_1, P_2, \varepsilon_1+\varepsilon_2, \varepsilon_1+\varepsilon_2)_{\text{avg}}$-multicast code. To this end, we first define $\tilde Z_1^n$ and $\tilde Z_2$ to be~$n$ independent copies of the standard normal random variable such that $\tilde Z_1^n$, $\tilde Z_2^n$ and $(X_1^n, X_2^n, Y_1^n, Y_2^n, Z_1^n, Z_2^n)$ are independent. In addition, there exist $w_1^*\in\mathcal{W}_1$ and $w_2^*\in \mathcal{W}_2$ such that
\begin{equation}
\Pr_{p_{W_1, W_2, Y_2^n}}\left\{\varphi_2(Y_2^n)\ne W_2|W_1=w_1^*\right\}= \arg \min_{w_1\in \mathcal{W}_1}\Pr_{p_{W_1, W_2, Y_2^n}}\left\{\varphi_2(Y_2^n)\ne W_2|W_1=w_1\right\}
\end{equation}
and
\begin{equation}
\Pr_{p_{W_1, W_2, Y_1^n}}\left\{\varphi_1(Y_1^n)\ne W_1|W_2=w_2^*\right\} = \arg \min_{w_2\in \mathcal{W}_2}\Pr_{p_{W_1, W_2, Y_1^n}}\left\{\varphi_1(Y_1^n)\ne W_1|W_2=w_2\right\},
\end{equation}
which implies from \eqref{eqn1LemmaIC} and \eqref{eqn2LemmaIC} that
\begin{equation}
 \Pr_{p_{W_2, Z_2^n}}\left\{\varphi_2(g_{21} f_1(w_1^*) + f_2(W_2)+Z_2^n)\ne W_2\right\}\le \varepsilon_2 \label{eqn3LemmaIC}
\end{equation}
and
\begin{equation}
\Pr_{p_{W_1,Z_1^n}}\left\{\varphi_1(f_1(W_1) + g_{12}f_2(w_2^*)+Z_1^n)\ne W_1\right\}\le \varepsilon_1\, . \label{eqn4LemmaIC}
\end{equation}
 Then, we define the stochastic decoders
\begin{align}
\varphi_1^\prime(Y_1^n)\triangleq \varphi_2\left(g_{21} f_1(w_1^*) +\frac{Y_1^n - f_1(\varphi_1(Y_1^n))}{g_{12}} + \sqrt{1-\frac{1}{g_{12}^2}}\tilde Z_2^n \right)  \label{eqn5LemmaIC}
\end{align}
and
\begin{align}
\varphi_2^\prime(Y_2^n)\triangleq \varphi_1\left( \frac{Y_2^n - f_2(\varphi_2(Y_2^n))}{g_{21}} + g_{12} f_2(w_2^*) + \sqrt{1-\frac{1}{g_{21}^2}}\tilde Z_1^n \right),  \label{eqn6LemmaIC}
\end{align}
where the randomness properties of the stochastic functions originate from not only~$\varphi_1$ and~$\varphi_2$ but also~$\tilde Z_1^n$ and~$\tilde Z_2^n$. Since
\begin{equation}
g_{21} f_1(w_1^*) + f_2(W_2)+Z_2^n
\end{equation}
and
\begin{equation}
g_{21} f_1(w_1^*) +\frac{Y_1^n - f_1(W_1)}{g_{12}} + \sqrt{1-\frac{1}{g_{12}^2}}\tilde Z_2^n
\end{equation}
have the same distribution by \eqref{channelLawIC},
it follows from \eqref{eqn3LemmaIC} and \eqref{eqn5LemmaIC} that
\begin{align}
& \Pr_{p_{W_1, W_2, Z_1^n, \tilde Z_2^n}}\left\{ \left\{\varphi_1^\prime(Y_1^n)\ne W_2\right\}\cap\left\{\varphi_1(Y_1^n)=W_1 \right\} \right\} \notag\\
& \quad \le \Pr_{p_{W_1, W_2, Z_1^n, \tilde Z_2^n}}\left\{ \varphi_2\left(g_{21} f_1(w_1^*) +\frac{Y_1^n - f_1(W_1)}{g_{12}} + \sqrt{1-\frac{1}{g_{12}^2}}\tilde Z_2^n\right)\ne W_2 \right\} \\
&\quad \le \varepsilon_2\, . \label{eqn7LemmaIC}
\end{align}
Combining \eqref{eqn5LemmaIC} and \eqref{eqn7LemmaIC}, we obtain
\begin{align}
&\Pr_{p_{W_1, W_2, Z_1^n, \tilde Z_2^n}}\left\{ \varphi_1(Y_1^n) \ne W_1  \text{ or } \varphi_1^\prime(Y_1^n) \ne W_2\right\}\notag\\
& \quad =\Pr_{p_{W_1, W_2, Z_1^n}}\left\{ \varphi_1(Y_1^n) \ne W_1  \right\}+\Pr_{p_{W_1, W_2, Z_1^n, \tilde Z_2^n}}\left\{ \left\{\varphi_1^\prime(Y_1^n)\ne W_2\right\}\cap\left\{\varphi_1(Y_1^n)=W_1 \right\} \right\}\\
&\quad \le \varepsilon_1 + \varepsilon_2\, .  \label{eqn8LemmaIC}
\end{align}
Following similar procedures for deriving \eqref{eqn8LemmaIC}, we obtain the following inequality by using  \eqref{channelLawIC}, \eqref{eqn4LemmaIC} and \eqref{eqn6LemmaIC}:
\begin{align}
\Pr_{p_{W_1, W_2, Z_1^n, \tilde Z_1^n}}\left\{ \varphi_2(Y_2^n) \ne W_2 \text{ or } \varphi_2^\prime(Y_2^n) \ne W_1\right\}\le \varepsilon_1 + \varepsilon_2\, .  \label{eqn9LemmaIC}
\end{align}
Replacing the decoding functions of the $(n, M_1^{(n)}, M_2^{(n)}, P_1, P_2, \varepsilon_1, \varepsilon_2)_{\text{avg}}$-code with $(\varphi_1, \varphi_1^\prime)$ and $(\varphi_2, \varphi_2^\prime)$ and keeping the encoding functions unchanged, we conclude from \eqref{eqn8LemmaIC} and \eqref{eqn9LemmaIC} that the resultant code is an $(n, M_1^{(n)}, M_2^{(n)}, P_1, P_2, \varepsilon_1+\varepsilon_2, \varepsilon_1+\varepsilon_2)_{\text{avg}}$-multicast code.
\end{IEEEproof}
\medskip

We are now ready to prove the strong converse theorem for the Gaussian IC under strong interference.
\medskip
\begin{IEEEproof}[Proof of Theorem~\ref{thmMainResultIC}]
Fix $\varepsilon_1>0$ and $\varepsilon_2>0$ such that
\begin{equation}
\varepsilon_1+\varepsilon_2<1. \label{eps1eps2LessThan1}
 \end{equation}
 As discussed at the beginning of Section~\ref{sectionIC}, it follows from Theorem~5.2 in \cite{Han81} that $\mathcal{C}_{0, 0}=\mathcal{R}_{\text{\tiny HK-S}}$ where the quantities $I_1$ and $I_2$ in $\mathcal{R}_{\text{\tiny HK-S}}$ are defined in \eqref{defI1} and \eqref{defI2} respectively. Since $\mathcal{C}_{0, 0} \subseteq \mathcal{C}_{\varepsilon_1, \varepsilon_2}$ for all non-negative real numbers $\varepsilon_1$ and $\varepsilon_2$ by definition,
  \begin{equation}
  \mathcal{R}_{\text{\tiny HK-S}}\subseteq \mathcal{C}_{\varepsilon_1, \varepsilon_2}. \label{knownInThmMainResultIC}
  \end{equation}
Therefore, it suffices to prove
\begin{equation}
\mathcal{C}_{\varepsilon_1, \varepsilon_2}\subseteq \mathcal{R}_{\text{\tiny HK-S}}. \label{goalInThmMainResultIC}
\end{equation}
To this end, fix a rate pair $(R_1, R_2)\in \mathcal{C}_{\varepsilon_1, \varepsilon_2}$. By definition, there exists a sequence of  $(n, M_1^{(n)}, M_2^{(n)}, P_1, P_2, \linebreak \varepsilon_1^{(n)}, \varepsilon_2^{(n)})_{\text{avg}}$-codes such that
\begin{equation}
\liminf_{n\rightarrow \infty}\frac{1}{n}\log M_i^{(n)} \ge R_i \label{eqn1ThmIC}
\end{equation}
and
\begin{equation}
\limsup_{n\rightarrow \infty}\varepsilon_i^{(n)} \le \varepsilon_i \label{eqn2ThmIC}
\end{equation}
for each $i\in\{1,2\}$. It then following from Lemma~\ref{lemmaIC} and \eqref{eqn2ThmIC} that there exists a sequence of  $(n, M_1^{(n)}, M_2^{(n)}, \linebreak P_1, P_2, \tilde \varepsilon_1^{(n)}, \tilde \varepsilon_2^{(n)})_{\text{avg}}$-multicast codes such that
\begin{equation}
\limsup_{n\rightarrow \infty}\tilde \varepsilon_i^{(n)} \le \varepsilon_1+\varepsilon_2 \label{eqn3ThmIC}
\end{equation}
for each $i\in\{1,2\}$.

Construct a subnetwork of the Gaussian IC formed by deleting~$\mathrm{d}_2$ as well as the links connecting to it. By inspection, the resultant subnetwork is a two-source Gaussian MAC and the sequence of $(n, M_1^{(n)}, M_2^{(n)}, P_1, P_2, \tilde \varepsilon_1^{(n)}, \tilde \varepsilon_2^{(n)})_{\text{avg}}$-multicast codes for the Gaussian IC induces a sequence of $(n, M_1^{(n)}, M_2^{(n)}, P_1, P_2,  \tilde \varepsilon_1^{(n)})_{\text{avg}}$-codes for the two-source Gaussian MAC. It then follows from \eqref{eqn1ThmIC} and \eqref{eqn3ThmIC} that $(R_1, R_2)$ is $(\varepsilon_1 + \varepsilon_2)$-achievable for the two-source Gaussian MAC, which implies from Theorem~\ref{thmMainResult}, \eqref{eps1eps2LessThan1} and \eqref{channelLawIC} that
\begin{align}
R_1 \le \frac{1}{2}\log\left(1+P_1\right), \label{eqn4ThmIC}
\end{align}
\begin{align}
R_2 \le \frac{1}{2}\log\left(1+g_{12}^2 P_2\right)
\end{align}
and
\begin{align}
R_1+R_2 \le \frac{1}{2}\log\left(1+P_1+g_{12}^2 P_2\right). \label{eqn5ThmIC}
\end{align}
Similarly, if we repeat the above procedures for the other two-source Gaussian MAC resulting from deleting $\mathrm{d}_1$ from the Gaussian IC, we obtain
\begin{align}
R_1 \le \frac{1}{2}\log\left(1+g_{21}^2 P_1\right),
\end{align}
\begin{align}
R_2 \le \frac{1}{2}\log\left(1+ P_2\right) \label{eqn6ThmIC}
\end{align}
and
\begin{align}
R_1+R_2 \le \frac{1}{2}\log\left(1+g_{21}^2 P_1+ P_2\right). \label{eqn7ThmIC}
\end{align}
Combining the bounds in \eqref{eqn4ThmIC}, \eqref{eqn5ThmIC}, \eqref{eqn6ThmIC}, \eqref{eqn7ThmIC}, the capacity region in \eqref{strong_interference}, and the strong interference conditions in \eqref{defI1} and \eqref{defI2}, we have $(R_1, R_2)\in  \mathcal{R}_{\text{\tiny HK-S}}$. Consequently, the outer bound in~\eqref{goalInThmMainResultIC} holds, and the theorem follows from \eqref{goalInThmMainResultIC} and the inner bound stated in~\eqref{knownInThmMainResultIC}.
\end{IEEEproof}
\appendix
\begin{IEEEproof}[Proof of Lemma~\ref{lemmaExpurgation}] Suppose an $(n, M_{\mathcal{I}}^{(n)}, P_\mathcal{I}, \varepsilon)_{\text{avg}}$-code is given for some $\varepsilon\in[0,1)$, and let
\begin{equation}
e_{w_\mathcal{I}}\triangleq\Pr\big\{\hat W_\mathcal{I}\ne w_\mathcal{I}\,\big|\, W_\mathcal{I}=w_\mathcal{I}\big\}
\end{equation}
be the probability of decoding error given that $w_\mathcal{I}$ is the message tuple transmitted by the sources. Then by choosing $w_\mathcal{I}$ one by one in an increasing order of $e_{w_\mathcal{I}}$, we can construct a set $\mathcal{D}\subseteq \mathcal{W}_\mathcal{I}$ such that
\begin{equation}
\Pr\big\{\hat W_\mathcal{I}\ne w_\mathcal{I} \, \big|\,W_\mathcal{I}=w_\mathcal{I}\big\}\le \frac{1+\varepsilon}{2} \label{eqn:prob_expurg}
\end{equation}
for all $w_\mathcal{I}\in \mathcal{D}$ and
\begin{equation}
|\mathcal{D}|\ge \left\lfloor \left(\frac{1-\varepsilon}{1+\varepsilon}\right)\prod_{i\in \mathcal{I}}M_i^{(n)} \right\rfloor. \label{boundOnAInProof}
 \end{equation}
 This is essentially an expurgation argument.
The bound in \eqref{eqn:prob_expurg} means that there exists an $\big(n, M_{\mathcal{I}}^{(n)}, P_\mathcal{I}, \mathcal{D}, \mathcal{I}, \frac{1+\varepsilon}{2}\big)_{\text{max}}$-code such that \eqref{boundOnAInProof} holds. Fix a nonempty $T\subseteq \mathcal{I}$. Define
 \begin{equation}
 \mathcal{D}_{w_{T^c}} \triangleq \{\tilde w_\mathcal{I} \in \mathcal{D}\, |\, \tilde w_{T^c}=w_{T^c}\} \label{defSetDT}
 \end{equation}
 for each $w_{T^c}\in \mathcal{W}_{T^c}$ such that
 \begin{equation}
 \sum_{w_{T^c}\in \mathcal{W}_{T^c}}|\mathcal{D}_{w_{T^c}}| = |\mathcal{D}|. \label{defSetDTsum}
 \end{equation}
 Since $|\mathcal{W}_{T^c}| = \prod_{i\in T^c}M_i^{(n)}$, it follows from \eqref{boundOnAInProof} and \eqref{defSetDTsum} that there exists a $w_{T^c}^*\in \mathcal{W}_{T^c}$ such that
 \begin{equation}
 | \mathcal{D}_{w_{T^c}^*}| \ge \left\lfloor \left(\frac{1-\varepsilon}{1+\varepsilon}\right)\prod_{i\in T}M_i^{(n)} \right\rfloor, \label{st4lemmaWringingInProof}
 \end{equation}
or otherwise we would obtain the following chain of inequalities which would eventually contradict \eqref{boundOnAInProof}:
 \begin{align}
 |\mathcal{D}|& \stackrel{\eqref{defSetDTsum}}{=} \sum_{w_{T^c}\in \mathcal{W}_{T^c}}|\mathcal{D}_{w_{T^c}}| \\*
 & < |\mathcal{W}_{T^c}|\left\lfloor \left(\frac{1-\varepsilon}{1+\varepsilon}\right)\prod_{i\in T}M_i^{(n)} \right\rfloor \\
 & = \prod_{i\in T^c}M_i^{(n)}\left\lfloor \left(\frac{1-\varepsilon}{1+\varepsilon}\right)\prod_{i\in T}M_i^{(n)} \right\rfloor \\*
 & \le \left\lfloor \left(\frac{1-\varepsilon}{1+\varepsilon}\right)\prod_{i\in \mathcal{I}}M_i^{(n)} \right\rfloor,
 \end{align}
 which contradicts \eqref{boundOnAInProof}. Due to \eqref{st4lemmaWringingInProof}, we can construct an $\big(n, M_{\mathcal{I}}^{(n)}, P_\mathcal{I}, \mathcal{D}_{w_{T^c}^*}, T, \frac{1+\varepsilon}{2}\big)_{\text{max}}$-code based on the $\big(n, M_{\mathcal{I}}^{(n)}, P_\mathcal{I}, \mathcal{D}, \mathcal{I}, \frac{1+\varepsilon}{2}\big)_{\text{max}}$-code such that they have the same message sets, encoding functions and decoding function and differ in only the support set of the message tuple $W_\mathcal{I}$ (cf.\ Definition~\ref{defCode}). In particular, the second statement in Definition~\ref{defCode} is satisfied because of the following reasons:
 \begin{enumerate}
 \item By construction, $W_\mathcal{I}$ is uniform on $\mathcal{D}_{w_{T^c}^*}$.
 \item For all $w_\mathcal{I}\in \mathcal{D}_{w_{T^c}^*}$, we have $w_{T^c}=w_{T^c}^*$ by \eqref{defSetDT}.
 \end{enumerate}
 Let $\mathcal{A}\triangleq \mathcal{D}_{w_{T^c}^*}$.
   It remains to show that \eqref{lemmaWringingSt} and \eqref{st2LemmaWringing} hold for the $\big(n, M_{\mathcal{I}}^{(n)}, P_\mathcal{I}, \mathcal{A}, T, \frac{1+\varepsilon}{2}\big)_{\text{max}}$-code. Recalling the definition of $\mathcal{A}_T$ in \eqref{defSetAT}, we obtain from \eqref{defSetDT} that
\begin{equation}
|\mathcal{A}|=|\mathcal{A}_T|=|\mathcal{D}_{w_{T^c}^*}|, \label{AT=DTInProof}
\end{equation}
which implies from \eqref{st4lemmaWringingInProof} that 
 \begin{align}
|\mathcal{A}|=|\mathcal{A}_T| \ge \left\lfloor \left(\frac{1-\varepsilon}{1+\varepsilon}\right)\prod_{i\in T}M_i^{(n)} \right\rfloor . \label{statement1InProof}
 \end{align}
 Consequently, \eqref{lemmaWringingSt} follows from \eqref{AT=DTInProof}, \eqref{statement1InProof} and \eqref{assumptionInLemmaWringing}.
 It remains to prove \eqref{st2LemmaWringing}. To this end, let
 $p_{W_\mathcal{I}, X_\mathcal{I}^n, Y^n, \hat W_T}$
  denote the probability distribution induced on the Gaussian MAC by the $\big(n, M_{\mathcal{I}}^{(n)}, P_\mathcal{I}, \mathcal{A}, T, \frac{1+\varepsilon}{2}\big)_{\text{max}}$-code, where
 \begin{equation}
 p_{W_T}(w_T)=\frac{1}{|\mathcal{A}_T| }\label{uniformDistributionATInProof}
 \end{equation}
 for all $w_T\in \mathcal{A}_T$ by Definition~\ref{defCode}. Using \eqref{uniformDistributionATInProof} and \eqref{lemmaWringingSt}, we obtain
 \begin{equation}
 p_{W_T}(w_T) \le \frac{1}{\prod_{i\in T}M_i^{(n)}}\cdot \left(\frac{2(1+\varepsilon)}{1-\varepsilon}\right)
 \end{equation}
 for each $w_T\in \mathcal{A}_T$.
 \end{IEEEproof}

\section*{Acknowledgements} The authors are extremely grateful to Yury Polyanskiy for pointing out an error in an earlier version of the manuscript. They would also like to thank the Associate Editor Prof.\ Sandeep Pradhan and the two anonymous reviewers for the useful comments which greatly improve the presentation of this paper.

The authors are supported by NUS grants R-263-000-A98-750/133, an MOE (Ministry of Education) Tier 2 grant R-263-000-B61-113, and the NUS Young Investigator Award R-263-000-B37-133.
\appendices

\ifCLASSOPTIONcaptionsoff
 \newpage
\fi


\begin{thebibliography}{10}
\providecommand{\url}[1]{#1}
\csname url@samestyle\endcsname
\providecommand{\newblock}{\relax}
\providecommand{\bibinfo}[2]{#2}
\providecommand{\BIBentrySTDinterwordspacing}{\spaceskip=0pt\relax}
\providecommand{\BIBentryALTinterwordstretchfactor}{4}
\providecommand{\BIBentryALTinterwordspacing}{\spaceskip=\fontdimen2\font plus
\BIBentryALTinterwordstretchfactor\fontdimen3\font minus
  \fontdimen4\font\relax}
\providecommand{\BIBforeignlanguage}[2]{{%
\expandafter\ifx\csname l@#1\endcsname\relax
\typeout{** WARNING: IEEEtran.bst: No hyphenation pattern has been}%
\typeout{** loaded for the language `#1'. Using the pattern for}%
\typeout{** the default language instead.}%
\else
\language=\csname l@#1\endcsname
\fi
#2}}
\providecommand{\BIBdecl}{\relax}
\BIBdecl

\bibitem{elgamal}
A.~{El~Gamal} and Y.-H. Kim, \emph{Network Information Theory}.\hskip 1em plus
  0.5em minus 0.4em\relax Cambridge, U.K.: Cambridge University Press, 2012.

\bibitem{ahl71}
R.~Ahlswede, ``{Multi-way communication channels},'' in \emph{Proc. IEEE
  Intl.~Symp.~on Inf.~Theory}, Thakadsor, Armenian SSR, 1971, pp. 23--52.

\bibitem{liao}
H.~H.~J. Liao, ``Multiple access channels,'' Ph.D. dissertation, University of
  Hawaii, Honolulu, 1972.

\bibitem{cover_75}
T.~M. Cover, ``Some advances in broadcast channels,'' in \emph{Advances in
  Communication Systems}, A.~J. Viterbi, Ed.\hskip 1em plus 0.5em minus
  0.4em\relax Academic Press, 1975, vol.~4, pp. 229--260.

\bibitem{wyner74}
A.~D. Wyner, ``{Recent results in the Shannon theory},'' \emph{IEEE Trans.~on
  Inf.~Theory}, vol.~20, pp. 2--10, 1974.

\bibitem{Wolfowitz}
J.~Wolfowitz, \emph{Coding Theorems of Information Theory}, 3rd~ed.\hskip 1em
  plus 0.5em minus 0.4em\relax Springer-Verlag, New York, 1978.

\bibitem{Tan_FnT}
V.~Y.~F. Tan, ``Asymptotic estimates in information theory with non-vanishing
  error probabilities,'' \emph{Foundations and Trends in Communications and
  Information Theory}, vol.~11, no. 1-2, pp. 1--183, 2014.

\bibitem{Scarlett15}
J.~Scarlett, A.~Martinez, and A.~{Guill\'en i F\`abregas}, ``Second-order rate
  region of constant-composition codes for the multiple-access channel,''
  \emph{IEEE Trans.~on Inf.~Theory}, vol.~61, no.~1, pp. 157--172, 2015.

\bibitem{Mol13}
E.~{MolavianJazi} and J.~N. Laneman, ``A second-order achievable rate region
  for {Gaussian} multi-access channels via a central limit theorem for
  functions,'' \emph{accepted to IEEE Trans.~on Inf.~Theory}, 2015.

\bibitem{TK14}
V.~Y.~F. Tan and O.~Kosut, ``On the dispersions of three network information
  theory problems,'' \emph{IEEE Trans.~on Inf.~Theory}, vol.~60, no.~2, pp.
  881--903, 2014.

\bibitem{Weng08}
L.~Weng, S.~S. Pradhan, and A.~Anastasopoulos, ``Error exponent regions for
  {Gaussian} broadcast and multiple-access channels,'' \emph{IEEE Trans.~on
  Inf.~Theory}, vol.~54, no.~7, pp. 2919--2942, 2008.

\bibitem{dueck81}
G.~Dueck, ``{The strong converse coding theorem for the multiple-access
  channel},'' \emph{{Journal of Combinatorics, Information \& System
  Sciences}}, vol.~6, no.~3, pp. 187--196, 1981.

\bibitem{Ahls76}
R.~Ahlswede, P.~G\'{a}cs, and J.~K\"{o}rner, ``{Bounds on conditional
  probabilities with applications in multi-user communication},'' \emph{Z.
  Wahrscheinlichkeitstheorie verw. Gebiete}, vol.~34, no.~3, pp. 157--177,
  1976.

\bibitem{Marton86}
K.~Marton, ``A simple proof of the blowing-up lemma,'' \emph{IEEE Trans.~on
  Inf.~Theory}, vol.~32, no.~3, pp. 445--446, 1986.

\bibitem{Csi97}
I.~Csisz\'{a}r and J.~{K\"{o}rner}, \emph{Information Theory: Coding Theorems
  for Discrete Memoryless Systems}.\hskip 1em plus 0.5em minus 0.4em\relax
  Cambridge University Press, 2011.

\bibitem{RagSason}
M.~Raginsky and I.~Sason, ``Concentration of measure inequalities in
  information theory, communications, and coding,'' \emph{Foundations and
  Trends in Communications and Information Theory}, vol.~10, no. 1-2, pp.
  1--246, 2013.

\bibitem{Ahl82}
R.~Ahlswede, ``{An elementary proof of the strong converse theorem for the
  multiple access channel},'' \emph{{Journal of Combinatorics, Information \&
  System Sciences}}, vol.~7, no.~3, pp. 216--230, 1982.

\bibitem{augustin}
U.~Augustin, ``Gedachtnisfreie kannale for diskrete zeit,''
  \emph{Z.~Wahrscheinlichkeitstheorie verw. Gebiete}, pp. 10--61, 1966.

\bibitem{Han98}
T.~S. Han, ``An information-spectrum approach to capacity theorems for the
  general multiple-access channel,'' \emph{IEEE Trans.~on Inf.~Theory},
  vol.~44, no.~7, pp. 2773--2795, 1998.

\bibitem{Han10}
------, \emph{Information-Spectrum Methods in Information Theory}.\hskip 1em
  plus 0.5em minus 0.4em\relax Springer Berlin Heidelberg, Feb 2003.

\bibitem{MolavianJaziThesis}
E.~MolavianJazi, ``A unified approach to {Gaussian} channels with finite
  blocklength,'' Ph.D. dissertation, University of Notre Dame, Jul. 2014.

\bibitem{PPV10}
Y.~Polyanskiy, H.~V. Poor, and S.~Verd\'{u}, ``Channel coding rate in the
  finite blocklength regime,'' \emph{IEEE Trans.~on Inf.~Theory}, vol.~56,
  no.~5, pp. 2307--2359, 2010.

\bibitem{wang09}
L.~Wang, R.~Colbeck, and R.~Renner, ``Simple channel coding bounds,'' in
  \emph{Proc. IEEE Intl.~Symp.~on Inf.~Theory}, Seoul, Korea, 2009, pp.
  1804--1808.

\bibitem{Han81}
T.~S. Han and K.~Kobayashi, ``{A new achievable rate region for the
  interference channel},'' \emph{IEEE Trans.~on Inf.~Theory}, vol.~27, no.~1,
  pp. 49--60, 1981.

\bibitem{Sato78}
H.~Sato, ``On the capacity region of a discrete two-user channel for strong
  interference,'' \emph{IEEE Trans.~on Inf.~Theory}, vol.~24, no.~3, pp.
  377--379, 1978.

\bibitem{LTM14}
S.-Q. Le, V.~Y.~F. Tan, and M.~Motani, ``A case where interference does not
  affect the channel dispersion,'' \emph{IEEE Trans.~on Inf.~Theory}, vol.~61,
  no.~5, pp. 2439--2453, 2015.

\end{thebibliography}
\end{document}